\documentstyle[12pt]{article}
\def\fileversion{v1.13}%
\def\filedate{6.5.93}%
\edef\epsfigRestoreAt{\catcode`@=\number\catcode`@\relax}%
\catcode`\@=11\relax
\immediate\write16{Document style option `epsfig', \fileversion\space
<\filedate> (edited by SPQR)}%
\newcount\EPS@Height
\newcount\EPS@Width
\newcount\EPS@xscale
\newcount\EPS@yscale
\def\psfigdriver#1{%
  \bgroup\edef\next{\def\noexpand\tempa{#1}}%
    \uppercase\expandafter{\next}%
    \def\LN{DVITOLN03}%
    \def\DVItoPS{DVITOPS}%
    \def\DVIPS{DVIPS}%
    \def\emTeX{EMTEX}%
    \def\OzTeX{OZTEX}%
    \def\Textures{TEXTURES}%
    \global\chardef\fig@driver=0
    \ifx\tempa\LN
        \global\chardef\fig@driver=0\fi
    \ifx\tempa\DVItoPS
        \global\chardef\fig@driver=1\fi
    \ifx\tempa\DVIPS
        \global\chardef\fig@driver=2\fi
    \ifx\tempa\emTeX
        \global\chardef\fig@driver=3\fi
    \ifx\tempa\OzTeX
        \global\chardef\fig@driver=4\fi
    \ifx\tempa\Textures
        \global\chardef\fig@driver=5\fi
  \egroup
\def\psfig@start{}%
\def\psfig@end{}%
\def\epsfig@gofer{}%
\ifcase\fig@driver
\typeout{WARNING! ****
 no specials for LN03 psfig}%
\or 
\def\psfig@start{}%
\def\psfig@end{\special{dvitops: import \@p@sfilefinal \space
\@p@swidth sp \space \@p@sheight sp \space fill}%
\if@clip \typeout{Clipping not supported}\fi
\if@angle \typeout{Rotating not supported}\fi
}%
\let\epsfig@gofer\psfig@end
\or 
\def\psfig@start{\special{ps::[begin]  \@p@swidth \space \@p@sheight \space%
        \@p@sbbllx \space \@p@sbblly \space%
        \@p@sbburx \space \@p@sbbury \space%
        startTexFig \space }%
        \if@angle
                \special {ps:: \@p@sangle \space rotate \space}
        \fi
        \if@clip
                \if@verbose
                        \typeout{(clipped to BB) }%
                \fi
                \special{ps:: doclip \space }%
        \fi
        \special{ps: plotfile \@p@sfilefinal \space }%
        \special{ps::[end] endTexFig \space }%
}%
\def\psfig@end{}%
\def\epsfig@gofer{\if@clip
                        \if@verbose
                           \typeout{(clipped to BB)}%
                        \fi
                        \epsfclipon
                  \fi
                  \epsfsetgraph{\@p@sfilefinal}%
}%
\or 
\typeout{WARNING. You must have a .bb info file with the Bounding Box
  of the pcx file}%
\def\psfig@start{}%
\def\psfig@end{\typeout{pcx import of \@p@sfilefinal}%
\if@clip \typeout{Clipping not supported}\fi
\if@angle \typeout{Rotating not supported}\fi
\raisebox{\@p@srheight true sp}{\special{em: graph \@p@sfilefinal}}}%
\def\epsfig@gofer{}%
\or 
\def\psfig@start{}%
\def\psfig@end{%
\EPS@Width\@p@swidth
\EPS@Height\@p@sheight
\divide\EPS@Width by 65781  
\divide\EPS@Height by 65781
\special{epsf=\@p@sfilefinal
\space
width=\the\EPS@Width
\space
height=\the\EPS@Height
}%
\if@clip \typeout{Clipping not supported}\fi
\if@angle \typeout{Rotating not supported}\fi
}%
\let\epsfig@gofer\psfig@end
\or 
\def\psfig@end{\if@clip
                        \if@verbose
                           \typeout{(clipped to BB)}%
                        \fi
                        \epsfclipon
                  \fi
\special{illustration \@p@sfilefinal\space scaled \the\EPS@xscale}%
}%
\def\psfig@start{}%
\let\epsfig\psfig
\else
\typeout{WARNING. *** unknown  driver - no psfig}%
\fi
}%
\newdimen\ps@dimcent
\ifx\undefined\fbox
\newdimen\fboxrule
\newdimen\fboxsep
\newdimen\ps@tempdima
\newbox\ps@tempboxa
\long\def\fbox#1{\leavevmode\setbox\ps@tempboxa\hbox{#1}\ps@tempdima\fboxrule
    \advance\ps@tempdima \fboxsep \advance\ps@tempdima \dp\ps@tempboxa
   \hbox{\lower \ps@tempdima\hbox
  {\vbox{\hrule height \fboxrule
          \hbox{\vrule width \fboxrule \hskip\fboxsep
          \vbox{\vskip\fboxsep \box\ps@tempboxa\vskip\fboxsep}\hskip
                 \fboxsep\vrule width \fboxrule}%
                 \hrule height \fboxrule}}}}%
\fi
\ifx\@ifundefined\undefined
\long\def\@ifundefined#1#2#3{\expandafter\ifx\csname
  #1\endcsname\relax#2\else#3\fi}%
\fi
\@ifundefined{typeout}%
{\gdef\typeout#1{\immediate\write\sixt@@n{#1}}}%
{\relax}%
\@ifundefined{epsfig}{}{\typeout{EPSFIG --- already loaded}\endinput}%
\@ifundefined{epsfbox}{\input epsf}{}%
\ifx\undefined\@latexerr
        \newlinechar`\^^J
        \def\@spaces{\space\space\space\space}%
        \def\@latexerr#1#2{%
        \edef\@tempc{#2}\expandafter\errhelp\expandafter{\@tempc}%
        \typeout{Error. \space see a manual for explanation.^^J
         \space\@spaces\@spaces\@spaces Type \space H <return> \space for
         immediate help.}\errmessage{#1}}%
\fi
\def\@whattodo{You tried to include a PostScript figure which
cannot be found^^JIf you press return to carry on anyway,^^J
The failed name will be printed in place of the figure.^^J
or type X to quit}%
\def\@whattodobb{You tried to include a PostScript figure which
has no^^Jbounding box, and you supplied none.^^J
If you press return to carry on anyway,^^J
The failed name will be printed in place of the figure.^^J
or type X to quit}%
\def\@nnil{\@nil}%
\def\@empty{}%
\def\@psdonoop#1\@@#2#3{}%
\def\@psdo#1:=#2\do#3{\edef\@psdotmp{#2}\ifx\@psdotmp\@empty \else
    \expandafter\@psdoloop#2,\@nil,\@nil\@@#1{#3}\fi}%
\def\@psdoloop#1,#2,#3\@@#4#5{\def#4{#1}\ifx #4\@nnil \else
       #5\def#4{#2}\ifx #4\@nnil \else#5\@ipsdoloop #3\@@#4{#5}\fi\fi}%
\def\@ipsdoloop#1,#2\@@#3#4{\def#3{#1}\ifx #3\@nnil
       \let\@nextwhile=\@psdonoop \else
      #4\relax\let\@nextwhile=\@ipsdoloop\fi\@nextwhile#2\@@#3{#4}}%
\def\@tpsdo#1:=#2\do#3{\xdef\@psdotmp{#2}\ifx\@psdotmp\@empty \else
    \@tpsdoloop#2\@nil\@nil\@@#1{#3}\fi}%
\def\@tpsdoloop#1#2\@@#3#4{\def#3{#1}\ifx #3\@nnil
       \let\@nextwhile=\@psdonoop \else
      #4\relax\let\@nextwhile=\@tpsdoloop\fi\@nextwhile#2\@@#3{#4}}%
\long\def\epsfaux#1#2:#3\\{\ifx#1\epsfpercent
   \def\testit{#2}\ifx\testit\epsfbblit
        \@atendfalse
        \epsf@atend #3 . \\%
        \if@atend
           \if@verbose
                \typeout{epsfig: found `(atend)'; continuing search}%
           \fi
        \else
                \epsfgrab #3 . . . \\%
                \epsffileokfalse\global\no@bbfalse
                \global\epsfbbfoundtrue
        \fi
   \fi\fi}%
\def\epsf@atendlit{(atend)}
\def\epsf@atend #1 #2 #3\\{%
   \def\epsf@tmp{#1}\ifx\epsf@tmp\empty
      \epsf@atend #2 #3 .\\\else
   \ifx\epsf@tmp\epsf@atendlit\@atendtrue\fi\fi}%

\chardef\trig@letter = 11
\chardef\other = 12

\newif\ifdebug 
\newif\ifc@mpute 
\newif\if@atend
\c@mputetrue 

\let\then = \relax
\def\r@dian{pt }%
\let\r@dians = \r@dian
\let\dimensionless@nit = \r@dian
\let\dimensionless@nits = \dimensionless@nit
\def\internal@nit{sp }%
\let\internal@nits = \internal@nit
\newif\ifstillc@nverging
\def \Mess@ge #1{\ifdebug \then \message {#1} \fi}%

{ 
        \catcode `\@ = \trig@letter
        \gdef \nodimen {\expandafter \n@dimen \the \dimen}%
        \gdef \term #1 #2 #3%
               {\edef \t@ {\the #1}
                \edef \t@@ {\expandafter \n@dimen \the #2\r@dian}%
                \t@rm {\t@} {\t@@} {#3}%
               }%
        \gdef \t@rm #1 #2 #3%
               {{%
                \count 0 = 0
                \dimen 0 = 1 \dimensionless@nit
                \dimen 2 = #2\relax
                \Mess@ge {Calculating term #1 of \nodimen 2}%
                \loop
                \ifnum  \count 0 < #1
                \then   \advance \count 0 by 1
                        \Mess@ge {Iteration \the \count 0 \space}%
                        \Multiply \dimen 0 by {\dimen 2}%
                        \Mess@ge {After multiplication, term = \nodimen 0}%
                        \Divide \dimen 0 by {\count 0}%
                        \Mess@ge {After division, term = \nodimen 0}%
                \repeat
                \Mess@ge {Final value for term #1 of
                                \nodimen 2 \space is \nodimen 0}%
                \xdef \Term {#3 = \nodimen 0 \r@dians}%
                \aftergroup \Term
               }}%
        \catcode `\p = \other
        \catcode `\t = \other
        \gdef \n@dimen #1pt{#1} 
}%

\def \Divide #1by #2{\divide #1 by #2} 

\def \Multiply #1by #2
       {{
        \count 0 = #1\relax
        \count 2 = #2\relax
        \count 4 = 65536
        \Mess@ge {Before scaling, count 0 = \the \count 0 \space and
                        count 2 = \the \count 2}%
        \ifnum  \count 0 > 32767 
        \then   \divide \count 0 by 4
                \divide \count 4 by 4
        \else   \ifnum  \count 0 < -32767
                \then   \divide \count 0 by 4
                        \divide \count 4 by 4
                \else
                \fi
        \fi
        \ifnum  \count 2 > 32767 
        \then   \divide \count 2 by 4
                \divide \count 4 by 4
        \else   \ifnum  \count 2 < -32767
                \then   \divide \count 2 by 4
                        \divide \count 4 by 4
                \else
                \fi
        \fi
        \multiply \count 0 by \count 2
        \divide \count 0 by \count 4
        \xdef \product {#1 = \the \count 0 \internal@nits}%
        \aftergroup \product
       }}%

\def\r@duce{\ifdim\dimen0 > 90\r@dian \then   
                \multiply\dimen0 by -1
                \advance\dimen0 by 180\r@dian
                \r@duce
            \else \ifdim\dimen0 < -90\r@dian \then  
                \advance\dimen0 by 360\r@dian
                \r@duce
                \fi
            \fi}%

\def\Sine#1%
       {{%
        \dimen 0 = #1 \r@dian
        \r@duce
        \ifdim\dimen0 = -90\r@dian \then
           \dimen4 = -1\r@dian
           \c@mputefalse
        \fi
        \ifdim\dimen0 = 90\r@dian \then
           \dimen4 = 1\r@dian
           \c@mputefalse
        \fi
        \ifdim\dimen0 = 0\r@dian \then
           \dimen4 = 0\r@dian
           \c@mputefalse
        \fi
        \ifc@mpute \then
                \divide\dimen0 by 180
                \dimen0=3.141592654\dimen0
                \dimen 2 = 3.1415926535897963\r@dian 
                \divide\dimen 2 by 2 
                \Mess@ge {Sin: calculating Sin of \nodimen 0}%
                \count 0 = 1 
                \dimen 2 = 1 \r@dian 
                \dimen 4 = 0 \r@dian 
                \loop
                        \ifnum  \dimen 2 = 0 
                        \then   \stillc@nvergingfalse
                        \else   \stillc@nvergingtrue
                        \fi
                        \ifstillc@nverging 
                        \then   \term {\count 0} {\dimen 0} {\dimen 2}%
                                \advance \count 0 by 2
                                \count 2 = \count 0
                                \divide \count 2 by 2
                                \ifodd  \count 2 
                                \then   \advance \dimen 4 by \dimen 2
                                \else   \advance \dimen 4 by -\dimen 2
                                \fi
                \repeat
        \fi
                        \xdef \sine {\nodimen 4}%
       }}%

\def\Cosine#1{\ifx\sine\UnDefined\edef\Savesine{\relax}\else
                             \edef\Savesine{\sine}\fi
        {\dimen0=#1\r@dian\multiply\dimen0 by -1
         \advance\dimen0 by 90\r@dian
         \Sine{\nodimen 0}%
         \xdef\cosine{\sine}%
         \xdef\sine{\Savesine}}}
\def\psdraft{\def\@psdraft{0}}%
\def\psfull{\def\@psdraft{1}}%
\psfull
\newif\if@scalefirst
\def\psscalefirst{\@scalefirsttrue}%
\def\psrotatefirst{\@scalefirstfalse}%
\psrotatefirst
\newif\if@draftbox
\def\psnodraftbox{\@draftboxfalse}%
\@draftboxtrue
\newif\if@noisy
\@noisyfalse
\newif\ifno@bb
\newif\if@bbllx
\newif\if@bblly
\newif\if@bburx
\newif\if@bbury
\newif\if@height
\newif\if@width
\newif\if@rheight
\newif\if@rwidth
\newif\if@angle
\newif\if@clip
\newif\if@verbose
\newif\if@prologfile
\def\@p@@sprolog#1{\@prologfiletrue\def\@prologfileval{#1}}%
\def\@p@@sclip#1{\@cliptrue}%
\newif\ifepsfig@dos  
\def\epsfigdos{\epsfig@dostrue}%
\epsfig@dosfalse
\newif\ifuse@psfig
\def\ParseName#1{\expandafter\@Parse#1}%
\def\@Parse#1.#2:{\gdef\BaseName{#1}\gdef\FileType{#2}}%

\def\@p@@sfile#1{%
\ifepsfig@dos
   \ParseName{#1:}%
\else
   \gdef\BaseName{#1}\gdef\FileType{}%
\fi
\def\@p@sfile{NO FILE: #1}%
\def\@p@sfilefinal{NO FILE: #1}%
        \openin1=#1
        \ifeof1\closein1
                \openin1=\BaseName.bb
                        \ifeof1\closein1
                                \if@bbllx\if@bblly\if@bburx\if@bbury
                                        \def\@p@sfile{#1}%
                                        \def\@p@sfilefinal{#1}%
                                        \fi\fi\fi
                                \else
                                        \@latexerr{ERROR.
PostScript file #1 not found}\@whattodo
                                        \@p@@sbbllx{100bp}%
                                        \@p@@sbblly{100bp}%
                                        \@p@@sbburx{200bp}%
                                        \@p@@sbbury{200bp}%
                                        \psdraft
                                \fi
                        \else
                                \closein1%
                                \edef\@p@sfile{\BaseName.bb}%
                                \typeout{using BB from \@p@sfile}%
                                \ifnum\fig@driver=3
                                  \edef\@p@sfilefinal{\BaseName.pcx}%
                                \else
                                 \ifepsfig@dos
                                 \edef\@p@sfilefinal{"`uncompress
                                   < \BaseName.Z"}%
                                \else
                                \edef\@p@sfilefinal{"`zcat `texfind
                                  #1.Z`"}%
                                \fi
                                \fi
                        \fi
        \else\closein1
                    \edef\@p@sfile{#1}%
                    \edef\@p@sfilefinal{#1}%
        \fi%
}%
\let\@p@@sfigure\@p@@sfile
\def\@p@@sbbllx#1{%
				            \@bbllxtrue
                \ps@dimcent=#1
                \edef\@p@sbbllx{\number\ps@dimcent}%
                \divide\ps@dimcent by65536
                \global\edef\epsfllx{\number\ps@dimcent}%
}%
\def\@p@@sbblly#1{%
                \@bbllytrue
                \ps@dimcent=#1
                \edef\@p@sbblly{\number\ps@dimcent}%
                \divide\ps@dimcent by65536
                \global\edef\epsflly{\number\ps@dimcent}%
}%
\def\@p@@sbburx#1{%
                \@bburxtrue
                \ps@dimcent=#1
                \edef\@p@sbburx{\number\ps@dimcent}%
                \divide\ps@dimcent by65536
                \global\edef\epsfurx{\number\ps@dimcent}%
}%
\def\@p@@sbbury#1{%
                \@bburytrue
                \ps@dimcent=#1
                \edef\@p@sbbury{\number\ps@dimcent}%
                \divide\ps@dimcent by65536
                \global\edef\epsfury{\number\ps@dimcent}%
}%
\def\@p@@sheight#1{%
                \@heighttrue
                \global\epsfysize=#1
                \ps@dimcent=#1
                \edef\@p@sheight{\number\ps@dimcent}%
}%
\def\@p@@swidth#1{%
                \@widthtrue
                \global\epsfxsize=#1
                \ps@dimcent=#1
                \edef\@p@swidth{\number\ps@dimcent}%
}%
\def\@p@@srheight#1{%
                \@rheighttrue\use@psfigtrue
                \ps@dimcent=#1
                \edef\@p@srheight{\number\ps@dimcent}%
}%
\def\@p@@srwidth#1{%
                \@rwidthtrue\use@psfigtrue
                \ps@dimcent=#1
                \edef\@p@srwidth{\number\ps@dimcent}%
}%
\def\@p@@sangle#1{%
                \use@psfigtrue
                \@angletrue
                \edef\@p@sangle{#1}%
}%
\def\@p@@ssilent#1{%
                \@verbosefalse
}%
\def\@p@@snoisy#1{%
                \@verbosetrue
}%
\def\@cs@name#1{\csname #1\endcsname}%
\def\@setparms#1=#2,{\@cs@name{@p@@s#1}{#2}}%
\def\ps@init@parms{%
                \@bbllxfalse \@bbllyfalse
                \@bburxfalse \@bburyfalse
                \@heightfalse \@widthfalse
                \@rheightfalse \@rwidthfalse
                \def\@p@sbbllx{}\def\@p@sbblly{}%
                \def\@p@sbburx{}\def\@p@sbbury{}%
                \def\@p@sheight{}\def\@p@swidth{}%
                \def\@p@srheight{}\def\@p@srwidth{}%
                \def\@p@sangle{0}%
                \def\@p@sfile{}%
                \use@psfigfalse
                \@prologfilefalse
                \def\@sc{}%
                \if@noisy
                        \@verbosetrue
                \else
                        \@verbosefalse
                \fi
                \@clipfalse
}%
\def\parse@ps@parms#1{%
                \@psdo\@psfiga:=#1\do
                   {\expandafter\@setparms\@psfiga,}%
\if@prologfile
\fi
}%
\def\bb@missing{%
        \if@verbose
            \typeout{psfig: searching \@p@sfile \space  for bounding box}%
        \fi
        \epsfgetbb{\@p@sfile}%
        \ifepsfbbfound
            \ps@dimcent=\epsfllx bp\edef\@p@sbbllx{\number\ps@dimcent}%
            \ps@dimcent=\epsflly bp\edef\@p@sbblly{\number\ps@dimcent}%
            \ps@dimcent=\epsfurx bp\edef\@p@sbburx{\number\ps@dimcent}%
            \ps@dimcent=\epsfury bp\edef\@p@sbbury{\number\ps@dimcent}%
        \else
            \epsfbbfoundfalse
        \fi
}
\newdimen\p@intvaluex
\newdimen\p@intvaluey
\def\rotate@#1#2{{\dimen0=#1 sp\dimen1=#2 sp
                  \global\p@intvaluex=\cosine\dimen0
                  \dimen3=\sine\dimen1
                  \global\advance\p@intvaluex by -\dimen3
                  \global\p@intvaluey=\sine\dimen0
                  \dimen3=\cosine\dimen1
                  \global\advance\p@intvaluey by \dimen3
                  }}%
\def\compute@bb{%
                \epsfbbfoundfalse
                \if@bbllx\epsfbbfoundtrue\fi
                \if@bblly\epsfbbfoundtrue\fi
                \if@bburx\epsfbbfoundtrue\fi
                \if@bbury\epsfbbfoundtrue\fi
                \ifepsfbbfound\else\bb@missing\fi
                \ifepsfbbfound\else
                \@latexerr{ERROR. cannot locate BoundingBox}\@whattodobb
                        \@p@@sbbllx{100bp}%
                        \@p@@sbblly{100bp}%
                        \@p@@sbburx{200bp}%
                        \@p@@sbbury{200bp}%
                        \no@bbtrue
                        \psdraft
                \fi
                \count203=\@p@sbburx
                \count204=\@p@sbbury
                \advance\count203 by -\@p@sbbllx
                \advance\count204 by -\@p@sbblly
                \edef\ps@bbw{\number\count203}%
                \edef\ps@bbh{\number\count204}%
                 \edef\@bbw{\number\count203}%
                \edef\@bbh{\number\count204}%
               \if@angle
                        \Sine{\@p@sangle}\Cosine{\@p@sangle}%

{\ps@dimcent=\maxdimen\xdef\r@p@sbbllx{\number\ps@dimcent}%

\xdef\r@p@sbblly{\number\ps@dimcent}%

\xdef\r@p@sbburx{-\number\ps@dimcent}%

\xdef\r@p@sbbury{-\number\ps@dimcent}}%
                        \def\minmaxtest{%
                           \ifnum\number\p@intvaluex<\r@p@sbbllx
                              \xdef\r@p@sbbllx{\number\p@intvaluex}\fi
                           \ifnum\number\p@intvaluex>\r@p@sbburx
                              \xdef\r@p@sbburx{\number\p@intvaluex}\fi
                           \ifnum\number\p@intvaluey<\r@p@sbblly
                              \xdef\r@p@sbblly{\number\p@intvaluey}\fi
                           \ifnum\number\p@intvaluey>\r@p@sbbury
                              \xdef\r@p@sbbury{\number\p@intvaluey}\fi
                           }%
                        \rotate@{\@p@sbbllx}{\@p@sbblly}%
                        \minmaxtest
                        \rotate@{\@p@sbbllx}{\@p@sbbury}%
                        \minmaxtest
                        \rotate@{\@p@sbburx}{\@p@sbblly}%
                        \minmaxtest
                        \rotate@{\@p@sbburx}{\@p@sbbury}%
                        \minmaxtest

\edef\@p@sbbllx{\r@p@sbbllx}\edef\@p@sbblly{\r@p@sbblly}%

\edef\@p@sbburx{\r@p@sbburx}\edef\@p@sbbury{\r@p@sbbury}%
                \fi
                \count203=\@p@sbburx
                \count204=\@p@sbbury
                \advance\count203 by -\@p@sbbllx
                \advance\count204 by -\@p@sbblly
                \edef\@bbw{\number\count203}%
                \edef\@bbh{\number\count204}%
}%
\def\in@hundreds#1#2#3{\count240=#2 \count241=#3
                     \count100=\count240        
                     \divide\count100 by \count241
                     \count101=\count100
                     \multiply\count101 by \count241
                     \advance\count240 by -\count101
                     \multiply\count240 by 10
                     \count101=\count240        
                     \divide\count101 by \count241
                     \count102=\count101
                     \multiply\count102 by \count241
                     \advance\count240 by -\count102
                     \multiply\count240 by 10
                     \count102=\count240        
                     \divide\count102 by \count241
                     \count200=#1\count205=0
                     \count201=\count200
                        \multiply\count201 by \count100
                        \advance\count205 by \count201
                     \count201=\count200
                        \divide\count201 by 10
                        \multiply\count201 by \count101
                        \advance\count205 by \count201
                     \count201=\count200
                        \divide\count201 by 100
                        \multiply\count201 by \count102
                        \advance\count205 by \count201
                     \edef\@result{\number\count205}%
}%
\def\compute@wfromh{%
                \in@hundreds{\@p@sheight}{\@bbw}{\@bbh}%
                \edef\@p@swidth{\@result}%
}%
\def\compute@hfromw{%
                \in@hundreds{\@p@swidth}{\@bbh}{\@bbw}%
                \edef\@p@sheight{\@result}%
}%
\def\compute@handw{%
                \if@height
                        \if@width
                        \else
                                \compute@wfromh
                        \fi
                \else
                        \if@width
                                \compute@hfromw
                        \else
                                \edef\@p@sheight{\@bbh}%
                                \edef\@p@swidth{\@bbw}%
                        \fi
                \fi
}%
\def\compute@resv{%
                \if@rheight \else \edef\@p@srheight{\@p@sheight} \fi
                \if@rwidth \else \edef\@p@srwidth{\@p@swidth} \fi
}%
\def\compute@sizes{%
        \if@scalefirst\if@angle
        \if@width
           \in@hundreds{\@p@swidth}{\@bbw}{\ps@bbw}%
           \edef\@p@swidth{\@result}%
        \fi
        \if@height
           \in@hundreds{\@p@sheight}{\@bbh}{\ps@bbh}%
           \edef\@p@sheight{\@result}%
        \fi
        \fi\fi
        \compute@handw
        \compute@resv
					           \EPS@Width=\@bbw
																\divide\EPS@Width by 1000
   												 \EPS@xscale=\@p@swidth \divide \EPS@xscale by \EPS@Width
					           \EPS@Height=\@bbh
																\divide\EPS@Height by 1000
   												 \EPS@yscale=\@p@sheight \divide \EPS@yscale by\EPS@Height
  \ifnum\EPS@xscale>\EPS@yscale\EPS@xscale=\EPS@yscale\fi
}
\def\psfig{\begingroup\@minisanitize\@@@psfig}
\def\epsfig{\begingroup\@minisanitize\@@@epsfig}

\def\@minisanitize{\@makeother\_\@makeother\:\@makeother\.\@makeother\$}

\def\@@@psfig#1{\vbox {%
        \ps@init@parms
        \parse@ps@parms{#1}%
        \ifnum\@psdraft=1
                \typeout{[\@p@sfilefinal]}%
                \if@verbose
                        \typeout{epsfig: using PSFIG macros}%
                \fi
                \psfig@method
        \else
                \epsfig@draft
        \fi
}
\endgroup
}%

\def\@@@epsfig#1{\vbox {%
        \ps@init@parms
        \parse@ps@parms{#1}%
        \ifnum\@psdraft=1
          \if@angle\use@psfigtrue\fi
          {\ifnum\fig@driver=1\global\use@psfigtrue\fi}%
          {\ifnum\fig@driver=3\global\use@psfigtrue\fi}%
          {\ifnum\fig@driver=4\global\use@psfigtrue\fi}%
          {\ifnum\fig@driver=5\global\use@psfigtrue\fi}%
                \ifuse@psfig
                        \if@verbose
                                \typeout{epsfig: using PSFIG macros}%
                        \fi
                        \psfig@method
                \else
                        \if@verbose
                                \typeout{epsfig: using EPSF macros}%
                        \fi
                        \epsf@method
                \fi
        \else
                \epsfig@draft
        \fi
}
\endgroup
}%

\def\epsf@method{%
        \epsfbbfoundfalse
        \if@bbllx\epsfbbfoundtrue\fi
        \if@bblly\epsfbbfoundtrue\fi
        \if@bburx\epsfbbfoundtrue\fi
        \if@bbury\epsfbbfoundtrue\fi
        \ifepsfbbfound\else\epsfgetbb{\@p@sfile}\fi
        \ifepsfbbfound
           \typeout{<\@p@sfilefinal>}%
           \epsfig@gofer
        \else
          \@latexerr{ERROR - Cannot locate BoundingBox}\@whattodobb
          \@p@@sbbllx{100bp}%
          \@p@@sbblly{100bp}%
          \@p@@sbburx{200bp}%
          \@p@@sbbury{200bp}%
                \count203=\@p@sbburx
                \count204=\@p@sbbury
                \advance\count203 by -\@p@sbbllx
                \advance\count204 by -\@p@sbblly
                \edef\@bbw{\number\count203}%
                \edef\@bbh{\number\count204}%
          \compute@sizes
          \epsfig@@draft
       \fi
}%
\def\psfig@method{%
        \compute@bb
        \ifepsfbbfound
          \compute@sizes
          \psfig@start
          \vbox to \@p@srheight true sp{\hbox to \@p@srwidth true
            sp{\hss}\vss\psfig@end}%
        \else
           \epsfig@draft
        \fi
}%
\def\epsfig@draft{\compute@bb\compute@sizes\epsfig@@draft}%
\def\epsfig@@draft{%
\typeout{<(draft only) \@p@sfilefinal>}%
\if@draftbox
        \hbox{\fbox{\vbox to \@p@srheight true sp{%
        \vss\hbox to \@p@srwidth true sp{ \hss
           {\tt\@p@sfilefinal}
                          \hss }\vss
        }}}%
\else
        \vbox to \@p@srheight true sp{%
        \vss\hbox to \@p@srwidth true sp{\hss}\vss}%
\fi
}%
\psfigdriver{dvips}%
\epsfigRestoreAt

\def\simlt{\stackrel{<}{{}_\sim}}
\def\simgt{\stackrel{>}{{}_\sim}}
\def\ov{\overline}

\def\msbu{m_{\tilde{b}_1}}
\def\msbd{m_{\tilde{b}_2}}
\def\mstu{m_{\tilde{t}_1}}
\def\mstd{m_{\tilde{t}_2}}

\def\mtl{m_{\tilde{t}_L}}
\def\mbl{m_{\tilde{b}_L}}
\def\mtr{m_{\tilde{t}_R}}
\def\mbr{m_{\tilde{b}_R}}
\def\otl{\ov{m}_{\tilde{t}_L}}
\def\otr{\ov{m}_{\tilde{t}_R}}
\def\obl{\ov{m}_{\tilde{b}_L}}
\def\omh{\ov{m}_h}
\def\omg{\ov{m}_{\chi}}
\def\st{\tilde{t}}
\def\stl{\tilde{t}_L}
\def\sbl{\tilde{b}_L}

\newcommand{\be}{\begin{equation}}
\newcommand{\ee}{\end{equation}}

\newcommand{\bear}{\begin{eqnarray}}
\newcommand{\eear}{\end{eqnarray}}

\def\R{v(T_c)/T_c}

\def\IJMPA #1 #2 #3 {Int.~J.~Mod.~Phys.~{\bf A#1}\ (19#2) #3}
\def\MPLA #1 #2 #3 {Mod.~Phys.~Lett.~{\bf A#1}\ (19#2) #3}
\def\NPB #1 #2 #3 {Nucl.~Phys.~{\bf B#1}\ (19#2) #3}
\def\PLB #1 #2 #3 {Phys.~Lett.~{\bf B#1}\ (19#2) #3}
\def\PR #1 #2 #3 {Phys.~Rep.~{\bf#1}\ (19#2) #3}
\def\PRD #1 #2 #3 {Phys.~Rev.~{\bf D#1}\ (19#2) #3}
\def\PTP #1 #2 #3 {Prog.~Theor.~Phys.~{\bf #1}\ (19#2) #3}
\def\PRL #1 #2 #3 {Phys.~Rev.~Lett.~{\bf#1}\ (19#2) #3}
\def\RMP #1 #2 #3 {Rev.~Mod.~Phys.~{\bf#1}\ (19#2) #3}
\def\ZPC #1 #2 #3 {Z.~Phys.~{\bf C#1}\ (19#2) #3}

\begin{document}

\begin{titlepage}

\title{\bf Dominant two-loop corrections to the MSSM finite temperature 
Effective Potential}

\author{
{\bf J.R. Espinosa} \thanks{Work supported by the Alexander-von-Humboldt 
Stiftung.} \\ Deutsches Elektronen Synchrotron DESY. \\
Notkestrasse 85.\ \ 22603 Hamburg. Germany}

\date{} 
\maketitle
\vspace{.5cm}
\def\baselinestretch{1.15}
\begin{abstract}
We show that two-loop corrections to the finite temperature effective 
potential in the MSSM can have a dramatic effect on the strength of the 
electroweak phase transition, making it more strongly first order.
The change in the order parameter $v/T_c$ can be as large as 75\%
of the one-loop daisy improved result. This effect can be decisive to
widen the region in parameter space where erasure of the created baryons 
by sphaleron processes after the transition is suppressed and 
hence, where electroweak baryogenesis might be successful. 
We find an allowed region with $\tan\beta\simlt 4.5$ and a 
Higgs boson with standard couplings and mass below $80\ GeV$ within the 
reach of LEP~II.
\end{abstract} 
\vspace{4cm}
\leftline{April 1996}

\thispagestyle{empty}

\vskip-20.cm
\rightline{{ DESY 96-064}}
\rightline{{ IEM--FT--122/96}}
\rightline{{ hep-ph/9604320}}
\vskip3in

\end{titlepage}

\def\baselinestretch{1.1}
\section{Introduction}

In recent years a considerable amount of work has 
been devoted to the study of 
the electroweak phase transition in the early Universe. A precise 
knowledge of it is crucial to determine if the very appealing idea of 
electroweak baryogenesis can be realized successfully at all (see 
\cite{ckn} for review and references). It is by now clear that the 
minimal Standard Model (SM) has serious problems to generate a sufficient 
amount of baryon asymmetry at the electroweak phase transition. The 
needed CP violation is far too small and in addition the phase transition 
is at most weakly first order for realistic values of the Higgs boson mass. 
That is, the ratio $v/T$, of the (temperature dependent) vacuum 
expectation value (vev) for the scalar field and the temperature, is 
significantly smaller than 1 at the moment of the transition. 
Then, erasure of the created $B+L$ asymmetry by unsuppressed 
sphaleron transitions can not be prevented.

Nevertheless, electroweak baryogenesis may still be successful in 
alternative models. A particularly well motivated extension of the SM is 
the Minimal Supersymmetric Standard Model (MSSM) \cite{MSSM}. It 
admits additional 
sources of CP violation \cite{CPSUSY} and the presence of many extra 
particles in its spectrum can influence significantly the details of the 
electroweak phase transition. A careful study of that transition in this 
model (see \cite{I,II} for a study of the temperature dependent one-loop 
daisy improved potential and \cite{previous} for previous studies) has 
shown that (if all experimental constraints 
are appropriately taken into account) a strong enough first order phase 
transition ($v/T\simgt 1$) can only take place in a very small region of 
parameter space. 
That region corresponds to large pseudoscalar mass, $m_A\gg m_Z$, small 
$\tan\beta$ ($\sim 2-3$), third generation squarks as light as otherwise 
allowed and a SM-like Higgs boson with mass just above the present 
experimental limit ($m_h\simgt 64 GeV$).

The purpose of this letter is to go beyond the one-loop daisy improved 
approximation for the finite temperature effective potential in the MSSM 
in order to see if the small window left for electroweak baryogenesis can 
be significantly widened. Similar studies, where the two-loop 
resummed potential is calculated, have been performed for the SM case 
\cite{TWOLSM} and sizeable corrections to different phase transition 
parameters were found. This in turn raised some doubts about the 
validity of perturbation theory and several alternative approaches were 
tried \cite{alter}.
Focusing on ${\cal R} \simeq v(T_c)/T_c$ ($T_c$ denotes the critical 
temperature, defined here by the coexistence of two degenerate minima in 
the potential), 
the quantity of interest for us, a careful comparison of two-loop resummed 
perturbative results and lattice simulations (see e.g. \cite{Jansen}) shows 
excellent agreement for moderate Higgs masses, indicating that 
perturbation theory is in better shape than previously thought. The 
inclusion of two-loop corrections tends to increase the value of 
$\R$ thus improving the prospects for preserving the baryon asymmetry 
against sphaleron erasure (although the improvement is not sufficient in the 
SM).

What can then be expected of a two-loop improvement for the MSSM? If the 
two-loop corrections turn out to be small, perturbation theory is trustworthy
but the window for baryogenesis will remain marginal. If on the other hand
very large corrections are found, perturbation theory will be suspect, and 
other techniques should be used. We hope to convince the reader that 
this is a naive expectation. We will show that large corrections indeed 
appear, making the phase transition much stronger, and nevertheless 
perturbation theory could be still considered reliable and under control.

Before embarking on all the details, let us consider briefly what are the 
relevant expansion parameters in the problem and roughly estimate the 
size of 
the corrections expected. In the Standard Model the order of the phase 
transition is 
determined by transverse gauge bosons through its contribution to the 
cubic term in 
the effective potential (see \cite{ckn} for details). The expansion 
parameter (in the minimum) 
can be identified as $\lambda/g^2$, where $\lambda$ is the quartic Higgs 
coupling and $g$ the SU(2) gauge coupling. The requirement 
$\lambda/g^2<1$ then puts an upper limit to the Higgs mass region where 
perturbation theory can be trusted: roughly $m_h^2\simlt m_W^2$. 

In the MSSM there are two main potential sources of differences with 
respect to the Standard Model: {\em i)} the presence of two Higgs doublets 
makes the effective potential a function of two background fields\footnote{
Even barring the possibility that other MSSM scalars take non 
zero vevs at high temperatures one should allow for the possibility of a 
non-zero relative phase between the two Higgs vevs. This case was studied 
in refs.~\cite{cpT}. The region of parameter space where it can be 
realized is disconnected with the one which will interest us here.} and 
{\em ii)} contributions of supersymmetric particles to the effective 
potential can have an important effect on the 
transition. Concerning {\em i)}, the two Higgs doublet potential is in fact 
tightly 
constrained by supersymmetry and it turns out that the best option for a 
strong transition corresponds to having only one light Higgs doublet (the 
case with a large pseudoscalar mass). Regarding the new 
contributions from supersymmetric particles it is easily realized that the 
key role is played by stops. If they are relatively light (of course, if 
all supersymmetric particles are heavy the Standard Model result for the 
corresponding Higgs mass is recovered) they drive the transition: being 
bosons, with a large number of degrees of freedom and sizeable coupling 
$(h_t)$ to the Higgs field, they dominate the cubic term in the potential 
even if their thermal masses are of order $g_sT$. 

In this situation the expansion 
parameter is $\lambda/h_t^2$ \cite{jrsintra}, where $h_t$ is the top 
Yukawa coupling. Therefore the scale to determine whether the Higgs mass is 
too heavy to trust perturbation theory is set now by the top mass $M_t$ 
while 
in the Standard Model it was set by $M_W$. This is just a naive estimate 
because the effect of the 
$g_sT$ screening can not be neglected and the coefficient in front of 
$\lambda/h_t^2$ that would give the proper expansion factor is unknown, 
but it suggests that perturbation 
theory can now be better behaved than it was in the SM for the same Higgs 
mass.

However, as we shall see, the situation in the MSSM is radically 
different from 
the SM case. It is not just a question of whether different effects will 
make 
the numerical coefficient in front of $\lambda/h_t^2$ large or small. The 
fact that squarks are {\em colored scalars} changes the picture 
completely: at two loops there are very important QCD corrections 
that affect $\R$ considerably and make the strong coupling $g_s$ to 
appear directly in the game. 

In the next section we review the one-loop daisy improved approximation 
for the MSSM potential \cite{I,II}. This will fix the notation and the 
interesting 
parameter range setting the stage for inclusion of two-loop dominant 
corrections, which will be the subject of section 3. 
Results and conclusion will be presented in sections 4 and 5 respectively. 
Some technical details are relegated to the appendices.  
\vspace{0.5cm}

\section{One-loop resummed potential} 

We assume from the beginning that the 
pseudoscalar mass $m_A$ is 
much larger than $M_Z$. In such case only one (combination) of the two 
Higgs 
doublets present in the MSSM is light. Then, the (light) physical 
Higgs boson has couplings to vector bosons
and fermions of SM strength, and in first approximation the 
experimental limit $m_h\geq 64\ GeV$ applies.
This simplifies the study of the 
potential which is then a function of just one background field as in the SM.
From supersymmetric particles only squarks of the third generation will 
be considered to be at the electroweak scale. This corresponds to the 
best situation regarding the strength of the electroweak phase transition 
(see \cite{I,II} for more details). Our model reduces then to the SM plus 
third generation squarks and a quartic Higgs coupling fixed by 
supersymmetry as shown below.

Working in 't~Hooft-Landau gauge and in 
$\ov{\mathrm MS}$-scheme\footnote{In refs. \cite{I,II} $\ov{\mathrm 
DR}$-scheme 
was used instead. We have checked that the numerical difference is 
irrelevant.}, we can write the finite temperature effective potential in 
the form
\be \label{total}
V_{\mathrm{eff}}(\varphi,T) = V_0(\varphi)
+ V_1(\varphi,0) + \Delta V_1(\varphi,T)
+\Delta V_{\mathrm{daisy}}(\varphi,T) \, ,
\ee
where
\bear
\label{v0}
V_0(\varphi) & = &  -\frac{1}{2}\mu^2\varphi^2 
+ \frac{1}{32} \varphi^4 (g^2+g'\,^2 )\cos^22\beta   \, ,
\\
\label{deltav}
V_1(\varphi,0) & = & \sum_i {n_i \over
64 \pi^2} m_i^4 (\varphi) \left[ \log {m_i^2 (\varphi) \over Q^2} - C_i \right]  
\, , \\
\label{deltavt}
\Delta V_1(\varphi,T) & = & {T^4 \over 2 \pi^2}  \sum_i
n_i \, J_i \left[ { m^2_i (\varphi) \over T^2 } \right]  \, ,
\\
\label{dvdaisy}
\Delta V_{\mathrm{daisy}}(\varphi,T) & = & - {T \over 12 \pi} \sum_i n_i
\left[ \ov{m}_i^3 (\varphi, T ) - m_i^3 (\varphi) \right] \, .
\eear

{\em 1)} The first contribution, eq.~(\ref{v0}), is just the tree level 
part. Note that the quartic Higgs coupling is fixed to be
\be 
\lambda=\frac{1}{8}(g^2+g'\,^2)\cos^22\beta.
\ee 
Here $\tan\beta$ is the ratio of the zero temperature vacuum expectation 
values of the two supersymmetric Higgses, $\langle 
H_2\rangle/\langle H_1 \rangle =v_2/v_1$, where $H_2$ is the doublet 
responsible for the masses of up-type quarks and $v_1^2+v_2^2=(246\ 
GeV)^2$. 

{\em 2)} The second term, 
eq.~(\ref{deltav}), is
the one-loop $T=0$ contribution. $Q$ is the renormalization scale,
where we choose for
definiteness $Q^2 = m_Z^2$. The $n_i$'s are the number of
degrees of freedom of the different species of particles, taken negative for 
fermions: \be
\label{multi}
n_t = - 12 \, ,  \;\;
n_{\st_1} = n_{\st_2} = 6 \,  , \;\;
n_W=6 \,  , \;\; n_Z=3  \,  , \;\; n_h=1 \,  , \;\; n_{\chi}=3 \, ,
\ee
and $C_i$ are constants equal to $5/6$ for vector bosons, and $3/2$ for
scalars and fermions.
The dominant piece of $V_1(\varphi,0)$ comes from
top and stop contributions.

Next, $m_i (\varphi)$ is the field-dependent mass of
the $i^{th}$ particle. The field-dependent top and bottom masses are
given by
\be
\label{tmass}
m_t^2(\varphi)=\frac{1}{2}h_t^2 \varphi^2 \sin^2\beta\, \;\;\; , \;\;\;
m_b^2(\varphi)=\frac{1}{2}h_b^2 \varphi^2 \cos^2\beta.
\ee
This implies the relations $h_{t,SM}=h_t\sin\beta$ and 
$h_{b,SM}=h_b\cos\beta$.

The entries of the field-dependent stop mass matrix are
\bear
\label{tlmass}
m_{\tilde{t}_L}^2 (\varphi) & = & m_{Q}^2 + m_t^2 (\varphi) +
D_{\tilde{t}_L}^2 (\varphi)  \, ,
\\
\label{trmass}
m_{\tilde{t}_R}^2 (\varphi) & = & m_{U}^2 + m_t^2 (\varphi) +
D_{\tilde{t}_R}^2 (\varphi)  \, ,
\eear
while the off-diagonal mixing between left and right-handed stops is 
given by \bear
m_{\tilde{t}_{LR}}^2 (\varphi) & = & \frac{h_t}{\sqrt{2}} 
(A_t \sin\beta + \mu \cos\beta) \varphi \equiv \frac{h_t}{\sqrt{2}}
X_t\varphi\, . 
\eear
Here $m_{Q}$, $m_{U}$ and $A_t$ are soft supersymmetry-breaking
mass
parameters\footnote{For constraints on these parameters we 
refer the reader to the discussion in \cite{I}.}, $\mu$ is a superpotential 
Higgs mass term, and \bear \label{dterms}
D_{\tilde{t}_L}^2(\varphi)& = & \frac{1}{4}\left( {1 \over 2}-
{2 \over 3}\sin^2 \theta_W \right)
(g^2 + g'\,^2 )\varphi^2\cos2\beta,\\
D_{\tilde{t}_R}^2(\varphi) &=&\frac{1}{4}\left( {2 \over 3}\sin^2 \theta_W 
\right) (g^2 + g'\,^2 )\varphi^2\cos2\beta,
\eear
are the $D$-term contributions to the stop mass matrix. The field-dependent 
stop masses,
$m^2_{\tilde{t}_{1,2}}$, are the eigenvalues of the matrix described.

For sbottoms one has
\bear
\label{blmass}
m_{\tilde{b}_L}^2 (\varphi) & = & m_{Q}^2 + m_b^2 (\varphi) +
D_{\tilde{b}_L}^2 (\varphi)  \, ,
\\
\label{brmass}
m_{\tilde{b}_R}^2 (\varphi) & = & m_{D}^2 + m_b^2 (\varphi) +
D_{\tilde{b}_R}^2 (\varphi)  \, ,
\eear
while the off-diagonal left-right mixing is now
\bear
m_{\tilde{b}_{LR}}^2 (\varphi) & = & \frac{h_b}{\sqrt{2}} (A_b \cos\beta + \mu 
\sin\beta) \varphi \equiv\frac{h_b}{\sqrt{2}}X_b\varphi\, . 
\eear
Here $m_{D}, A_b$ are soft supersymmetry-breaking
masses, and
\bear
\label{dbterms}
D_{\tilde{b}_L}^2(\varphi)& = & \frac{1}{4}\left( -{1 \over 2}+
{1 \over 3}\sin^2 \theta_W \right)
(g^2 + g'\,^2 )\varphi^2\cos2\beta,\\
D_{\tilde{b}_R}^2(\varphi) &=&\frac{1}{4}\left( -{1 \over 3}\sin^2 \theta_W 
\right) (g^2 + g'\,^2 )\varphi^2\cos2\beta\, .
\eear
The field-dependent masses for the gauge bosons 
are given by
\be
\label{gauge}
m_W^2 (\varphi) = \frac{1}{4} g^2 \varphi^2 \, ,
\;\;\;\;\;\;
m_Z^2 (\varphi) = \frac{1}{4}(g^2 + g'\,^2 ) \varphi^2 \, ,
\ee
and for Higgs and Goldstone scalars by
\be
\label{scalars}
m_h^2=3\lambda\varphi^2-\mu^2\, ,
\;\;\;\;\;\;
m^2_{\chi_3^0}=m^2_{\chi^{\pm}}=\lambda\varphi^2-\mu^2.
\ee

{\em 3)} The third term in the potential, eq.~(\ref{deltavt}), is the
one-loop contribution due to temperature effects. Here $J_i=J_+ (J_-)$ if the
$i^{th}$
particle is a boson (fermion), is the free energy of an ideal gas of 
particles of mass $m_i(\varphi)$,
\be
\label{ypsilon}
J_{\pm} (y^2)=J_{\pm} (m^2/T^2) \equiv \int_0^{\infty} dx \, x^2 \,
\log \left( 1 \mp e^{- \sqrt{x^2 + y^2}} \right) \, .
\ee

{\em 4)} The last term, eq.~(\ref{dvdaisy}), is a
correction
coming from the resummation of the leading infrared-dominated
higher-loop
contributions, associated with the so-called daisy diagrams. The sum
runs over bosons only. The effect of this contribution is to replace the 
pure cubic term $m^3(\varphi)$ that arises from 
(\ref{deltavt}) by $\ov{m}^3(\varphi)$. The masses $\ov{m}_i^2 (\varphi,T)$ 
are 
obtained from the ${m}_i^2 (\varphi)$ by adding\footnote{In general, 
$\ov{m}_i^2$ will be the eigenvalues of the corresponding mass matrix 
after addition of thermal contributions.} the leading $T$-dependent 
self-energy contributions, which are proportional to $T^2$. We recall 
that, in the gauge
boson sector, only the longitudinal components ($W_L, Z_L, \gamma_L$)
receive
such contributions\footnote{Longitudinal photons do contribute to the 
effective potential due to the effect described in the previous footnote.}. 
Also note \be \label{multiL}
n_{W_L}=2 \, ,  \;\; n_{Z_L}=n_{\gamma_L}=1 \, .
\ee
The (leading order) thermal polarizations for different particles can be 
found 
in ref.~\cite{I} and are reproduced in appendix A together with some 
extra polarizations needed for two-loop corrections to the potential. 
We assume here that only 
SM particles and third generation squarks contribute to these thermal
masses, the rest being heavy and decoupled. As explained in \cite{I} this 
is the best case for a strong first order transition and in practice
corresponds to the case of heavy gluinos, so that the results found can 
be interpreted as the best guess for a realistic supersymmetric spectrum.

The contribution of stops to $\Delta V_{daisy}$ is responsible 
for the 
enhancement of ${\cal R}=v(T_c)/T_c$ with respect to the SM result. The 
final 
region in parameter space where ${\cal R}>1$ is determined by the interplay 
between different opposite effects: soft and thermal screening masses 
tend to decrease ${\cal R}$ by screening the pure cubic behaviour of 
$m_{\tilde t}^3(\varphi)$ while a large Yukawa coupling (thus a large top 
mass) tend to increase it. For a given value of $h_t$ (and $\tan\beta$) 
there is a minimum 
allowed value of $m_{Q}$ [the constraint of $\Delta\rho$ on the 
mass splitting of the $(\tilde{t},\tilde{b})_L$ doublet gets weaker for 
larger $m_Q$] and this 
will correspond to the best case. To determine which of the competing 
effects is stronger one has to study numerically the potential as was 
done in \cite{I,II}. The main conclusion of those papers is that in some 
regions of parameter space the cubic term from stops is effective and 
actually dominates the electroweak phase transition.   

Regions of ${\cal R}>1$ can be found for $135\ GeV\simlt M_t\simlt 170\ 
GeV$. The best case always corresponds to negligible mixing
in the stop mass matrix ($m_{\tilde{t}_{LR}}^2\sim 0$), low $\tan\beta$ 
($\sim 2-3$), a Higgs 
mass close to the experimental limit ($m_h\leq 70\ GeV$) and squarks of 
the third generation as light as allowed by the $\Delta\rho$ constraint.

To fix ideas, for 
$m_t=150\ GeV$ (which corresponds to a physical pole mass $M_t=156\ GeV$),
$m_{Q}=70\ GeV$ [this gives $\Delta\rho(t,b)+\Delta\rho(\stl,\sbl)<0.01$],
$m_{U}=m_{\tilde{t}_{LR}}=0$ and $\tan\beta=2.5$,  $v(T_c)/T_c=1.02$
and so this point in parameter 
space corresponds roughly to the border of the ${\cal R}>1$ area.
The Higgs mass is $\sim 65\ GeV$. \vspace{0.5cm}

\section{Two-loop resummed potential} 

The complete two-loop resummed 
potential for the contributions 
of SM particles can be found in refs. \cite{TWOLSM}. Regarding its effect
on the quantity ${\cal R}=v(T_c)/T_c$, the most important corrections are 
those 
coming from bosonic setting sun diagrams, which cause logarithms of 
masses to appear. The importance of these contributions was first pointed 
out by Bagnasco and Dine in \cite{TWOLSM}. In particular, the most 
important effect is due to logarithms of transverse vector boson masses, 
unscreened at leading order. It is enough for our purpose to keep only 
logarithmic SM terms. We will also set $g'=0$ for all two-loop 
corrections. This will simplify the analytic results being numerically 
an excellent approximation. In addition we will use throughout a high 
temperature expansion keeping contributions to the effective potential up 
to order $g^4$ (where for power counting $g$ represents gauge, Yukawa 
couplings or $\sqrt{\lambda}$). 
In our formulas, $M$ will represent the (field-dependent) 
weak vector boson mass and  $M_L$ the longitudinal, Debye-screened mass 
(see appendix A): 
\be
M^2=\frac{1}{4}g^2\varphi^2\; ,\;\;\; 
M_L^2=M^2+\frac{7}{3}g^2T^2.
\ee
Then, the dominant two-loop resummed SM piece of the potential can be 
written as
\be
\label{v2sm}
V^{(2)}_{SM}=\frac{g^2}{16\pi^2}T^2\left[M^2\left(\frac{3}{4}\log\frac{M_L}{T}-
\frac{51}{8}\log\frac{M}{T}\right)+\frac{3}{2}(M^2-4M_L^2)\log\frac{M+2M_L}{3T}
\right].
\ee
Some comments are in order. Extra logarithmic contributions 
involving scalar (Higgs) masses  have been systematically neglected 
(although these were also kept for the numerical analysis). 
That is a good approximation because near the critical temperature their
effective mass is small. We have kept logarithms of longitudinal vector 
boson masses [note that only the $\log(M/T)$ was kept in Bagnasco and Dine 
paper]. They are not important numerically because the screening of gauge 
bosons is very effective here but we keep them for consistency: similar 
logarithms will appear for the supersymmetric contribution. Terms like the 
last one in (\ref{v2sm}) have a linear dependence
on $\varphi$ for small field. As is well known, this linear dependence 
cancels when all resummed two-loop contributions are kept. We shall then
add the terms needed to insure this cancellation here. These are 
\cite{TWOLSM}
\be
\delta V^{(2)}_{SM}=\frac{3g^2}{16\pi^2}MM_LT^2.
\ee

There are also non logarithmic contributions involving  $g_s$ and 
$h_t$ that can be sizeable in principle. They are \cite{TWOLSM}:
\be
\delta' V^{(2)}_{SM}=\frac{m_t^2(\varphi)T^2}{64\pi^2}\left[
16 g_s^2 \left(\frac{8}{3}\log2-\frac{1}{2}-c_B\right)+
9 h_{t,SM}^2 \left(\frac{4}{3}\log2-c_B\right)
\right].
\ee
Here, the constant $c_B$ is given by
\be
c_B=\log(4\pi)-\gamma_E
\ee
where $\gamma_E$ is Euler's constant. This correction affects mainly the 
numerical value of the critical temperature while its effect on ${\cal R}$ 
is very small.
 
\begin{figure}[hbt]
\centerline{
\psfig{figure=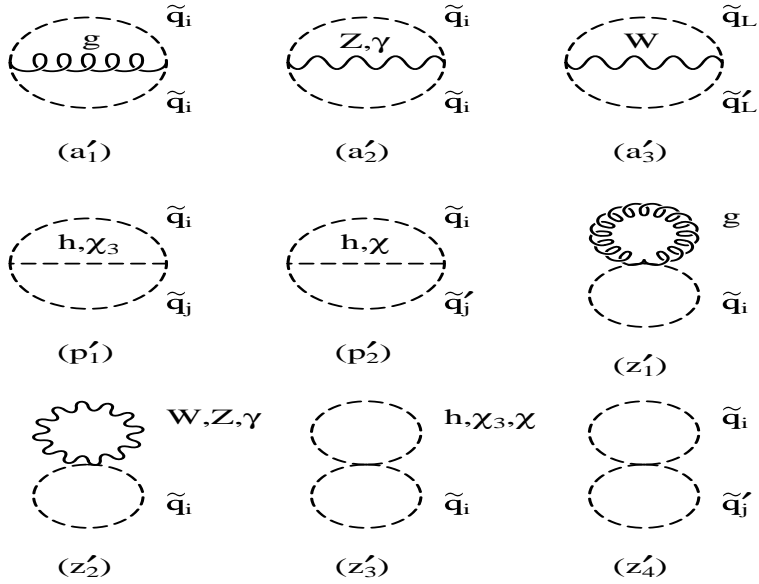,height=8cm,width=8cm,bbllx=6.cm,bblly=3.cm,bburx=17.cm,bbury=22cm}}
\caption{\footnotesize Two-loop graphs involving third generation squarks.}
\end{figure}

The extra supersymmetric diagrams\footnote{For MSSM Feynman rules see 
e.g. \cite{Rosiek}.} we have to consider are depicted in figure 1. Note 
that, with our assumptions on the supersymmetric spectrum,  R-parity 
conservation implies that squarks appear always in 
closed loops. Furthermore, there are no fermionic diagrams because all 
${\mathrm R}=-1$ fermions are assumed heavy (two-loop diagrams with e.g. a 
top 
quark and a stop involve also higgsinos or gluinos to close the fermion 
loop). We label the diagrams in a similar way as in refs. \cite{TWOLSM}.
By $\tilde{q}_i$ we represent a squark of a given flavour $q$ and chirality 
$i=L,R$. Then, $\tilde{q}'_j$ would stand for a squark of different 
flavour and different chirality. 

The dominant logarithmic contribution (plus linear terms) can then be 
written, in the high temperature expansion (see appendix B for the integral 
expression), as the sum of the following pieces ($N_c=3$ is the number of 
colours): \bear \label{va1}
V_{(a'_1)}&=&-\frac{g_s^2(N_c^2-1) 
T^2}{16\pi^2}\left[\otl^2\log\frac{2\otl}{3T} 
+\otr^2\log\frac{2\otr}{3T}+\obl^2\log\frac{2\obl}{3T} 
 \right],
\eear
\bear
\label{va23}
V_{(a'_{2,3})}&=&\frac{g^2N_c T^2}{32\pi^2}\left[
-\frac{3M}{2}(\otl+\obl)\right.\nonumber\\ \vspace{4mm}
&+&\frac{1}{4}(M^2-4\otl^2)
\log\frac{M+2\otl}{3T} 
+\frac{1}{4}(M^2-4\obl^2)
\log\frac{M+2\obl}{3T}\\ \vspace{4mm}
&+&\left.[M^2-2(\otl^2+\obl^2)]
\log\frac{M+\otl+\obl}{3T}+\frac{(\otl^2-\obl^2)^2}{M^2}
\log\frac{M+\otl+\obl}{\otl+\obl} 
\right]\nonumber,
\eear
\bear
\label{vp12}
V_{(p'_{1,2})}&=&\frac{N_c \varphi^2 T^2}{32\pi^2}
\left[\left(h_{t,SM}^2+\frac{1}{4}g^2\cos 2\beta\right)^2
\log\frac{\omh+2\otl}{3T}
+\left(h_{t,SM}^2\right)^2
\log\frac{\omh+2\otr}{3T}\right.\\ \vspace{4mm}
&+&\left.\left(h_{t,SM}^2+\frac{1}{2}g^2\cos 2\beta\right)^2
\log\frac{\omg+\otl+\obl}{3T}+\left(\frac{1}{4}g^2\cos 2\beta\right)^2
\log\frac{\omh+2\obl}{3T}\right]\nonumber .
\eear
In this last contribution we are neglecting stop mixing 
($m_{\tilde{t}_{LR}}\sim 0$). As was shown in \cite{I,II} for the one-loop 
potential, this is the best case to maximize ${\cal R}$. The two-loop
resummed potential for non-zero stop mixing can be found in appendix C.
 
The $(z')$ diagrams do not contribute logarithmic terms but they are 
needed to cancel the linear dependence present in the 
previous pieces:
\bear
\label{vz}
V_{(z')}&=&\frac{3g^2}{32\pi^2}N_c T^2M(\otl+\obl).
\eear

From the rest of non logarithmic terms we will keep those involving 
$g_s^2$ and $h_t^2$ only:
\bear
\label{vlin}
\delta V_{(a'_1)}&=&-\frac{g_s^2T^2}{64\pi^2}(N_c^2-1)(c_2-1)(\otl^2+\otr^2),
\nonumber\\ \vspace{4mm}
\delta V_{(p'_{1,2})}&=&\frac{3N_c}{128\pi^2}T^2h_{t,SM}^4 c_2 
\varphi^2 ,\nonumber\\ \vspace{4mm}
\delta 
V_{(z'_{1,2})}&=&\frac{g_s^2T^2}{32\pi^2}(N_c^2-1)\left[
\Pi_{gL}^{1/2}(\otl+\otr)+
\frac{1}{6}(\otl^2+\otr^2)\right]+\delta_{c_B} V_{(z'_{1,2})},
\nonumber\\ \vspace{4mm}
\delta 
V_{(z'_{3,4})}&=&\frac{N_c 
T^2}{16\pi^2}\left[\frac{g_s^2}{6}(N_c+1)(\otl^2+ \otr^2)
+ h_t^2\otr(\otl+\obl)
\right.\nonumber\\
&&+\left.\frac{1}{2}h_{t,SM}^2(\omh(\otl+\otr)+\omg(\otl+3\otr))
\right]+\delta_{c_B} V_{(z'_{3,4})}, 
\eear
where $c_2\simeq 3.3025$. Here $\Pi_{gL}$ is the Debye mass for 
longitudinal gluons as given in appendix~A. We have neglected 
throughout the bottom Yukawa coupling, which is small for small $\tan\beta$.
Then terms dependent on $\ov{m}_{{\tilde b}_R}$ are not shown.

The complete terms $\delta_{c_B}V$ add up to give
\be
\label{vcb}
\frac{c_B}{16\pi^2}\left[6\sum_im^2_{\tilde{q}_i}\Pi^{SUSY}_{\tilde{q}_i}
+m_{h}^2\Pi^{SUSY}_{h}+3m_{\chi}^2\Pi^{SUSY}_{\chi}\right].
\ee
As expected, this term plus a similar term from Standard Model 
polarizations, combines with the one-loop unresummed scalar contribution
\be
\frac{c_B}{32\pi^2}\sum_in_im_i^4,
\ee
to give the same result but with $m_i\rightarrow\overline{m}_i$ (a welcome 
check of the calculation). In (\ref{vcb}) $\Pi_i^{SUSY}$ is the full
thermal mass for squarks while for the Higgs modes it includes only
the contribution of supersymmetric particles (see appendix A):
\be
\Pi_{h}^{SUSY}=\Pi_{\chi}^{SUSY}=\frac{1}{2}h_t^2\sin^2\beta\, T^2.
\ee

\begin{figure}[hbt]
\centerline{
\psfig{figure=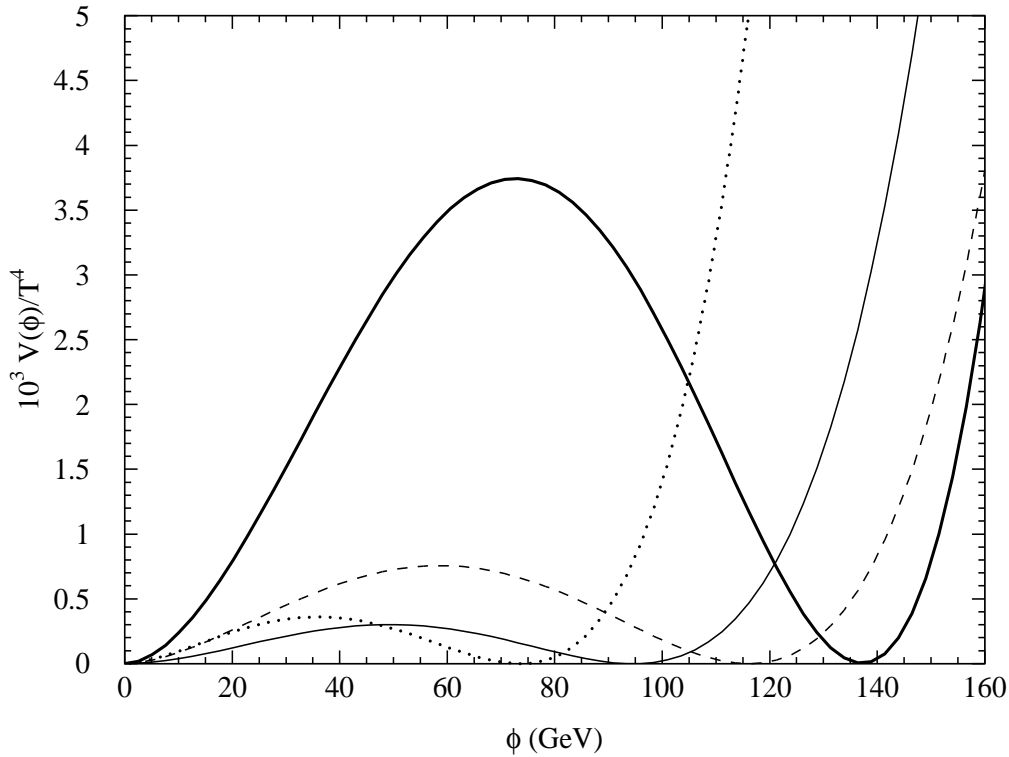,angle=-90,height=10cm,width=12cm,bbllx=2cm,bblly=4cm,bburx=19cm,bbury=25cm}}
\caption{\footnotesize Scaled potential at the critical temperature for 
different 
approximations: one-loop resummed potential (thin solid line); potential 
with two-loop resummed contributions from SM particles only (dashed); with
two-loop resummed supersymmetric non-QCD contributions added (dotted); with
all dominant two-loop resummed contributions (thick solid). [$M_t=156\ 
GeV$, $m_Q=70\ GeV$, $m_U=m_{{\tilde t}_{LR}}=0$ and $\tan\beta=2.5$].} 
\end{figure} 

\section{Results} 
In figure 2, different approximations to the scaled effective potential 
[$V(\varphi)/T^4$] are plotted at the corresponding critical 
temperature.  
Parameters are fixed as in the example at the end of section 2:
top pole mass $M_t=156\ GeV$,
$m_{Q}=70\ GeV$, $m_{U}=m_{\tilde{t}_{LR}}=0$ and $\tan\beta=2.5$.
The thin 
solid line corresponds to the one-loop resummed potential for 
which ${\cal R}=1.02$.
The effect of including two-loop corrections from Standard Model particles
is shown by the dashed line: the transition becomes stronger and ${\cal R}$ 
increases
up to 1.20. This effect can be traced back \cite{TWOLSM} to the presence of 
corrections
of the form $-\varphi^2\log\varphi$ coming from transverse gauge boson modes
[see eq.~(\ref{v2sm})].
A contribution of this form has a direct effect shifting the value of 
the vev at the critical temperature. The minus sign in front implies 
that the shift is towards larger values.  
To examine two-loop corrections from supersymmetric particles it is 
convenient to split them in two categories: QCD contributions and the rest.
Looking first at non QCD terms, the dominant effect [see 
eqs.~(\ref{va1}-\ref{vlin})]
is of the form $+\varphi^2\log\ov{\varphi}$ 
(where $\ov{\varphi}$ represents some 
screened masses). The net positive sign indicates that these corrections 
contribute to make the phase transition weaker. This effect can be seen in 
figure 2 where the dotted line includes non QCD squark contributions. 
The corresponding ${\cal R}$ falls now down to 0.97, slightly below the 
one-loop resummed result.
Then, disregarding QCD corrections, the convergence of the perturbative  
series for ${\cal R}$ seems to be much better in the supersymmetric case, 
when the 
phase transition is driven by squarks, than in the Standard Model (for 
the same Higgs mass).

The inclusion of QCD corrections changes completely this picture: a look at
(\ref{va1}) reveals that these corrections are of the form 
$-\varphi^2\log\ov{\varphi}$ and sizeable, implying that ${\cal R}$ will be 
shifted to 
larger values. The increase in ${\cal R}$ can be as dramatic as shown in 
figure 2.
The thick solid line gives the scaled potential with all dominant 
corrections included. The phase transition in this approximation is much 
stronger and ${\cal R}=1.75$ (a correction of $75\%$ over the one-loop 
result!).

This huge effect makes perturbative validity dubious. Nevertheless, one 
should keep 
in mind that the effect is very large because QCD corrections appear only 
at two loops (the same happens in the Standard Model but there, 
the effect is associated with a fermionic loop and so the final effect on 
${\cal R}$ is much less important) and not at one-loop\footnote{This 
situation is by no 
means unique and appears in many other cases. Similar examples can be 
found already in the MSSM Higgs sector, e.g. when considering radiative 
corrections to the mass of the lightest scalar \cite{Radmh} 
or QCD radiative corrections to hadronic Higgs decays \cite{Spira}.}. 
Then, a large ratio of 
the two-loop to the one-loop result does not necessarily imply that 
perturbation theory is not applicable. To settle definitively this issue 
one should estimate the size of three-loop corrections. However, that is 
beyond the aim of this paper. Note also that lattice simulations would 
face here the problem that squarks carry color so that $SU(3)_c$ can no 
longer be ignored as in the SM once fermions were integrated out. 
Lacking an estimate of higher order corrections, 
certainly this two-loop result indicates that the transition is probably 
much stronger than what the one-loop resummed approximation indicated.

\begin{figure}[hbt]
\centerline{
\psfig{figure=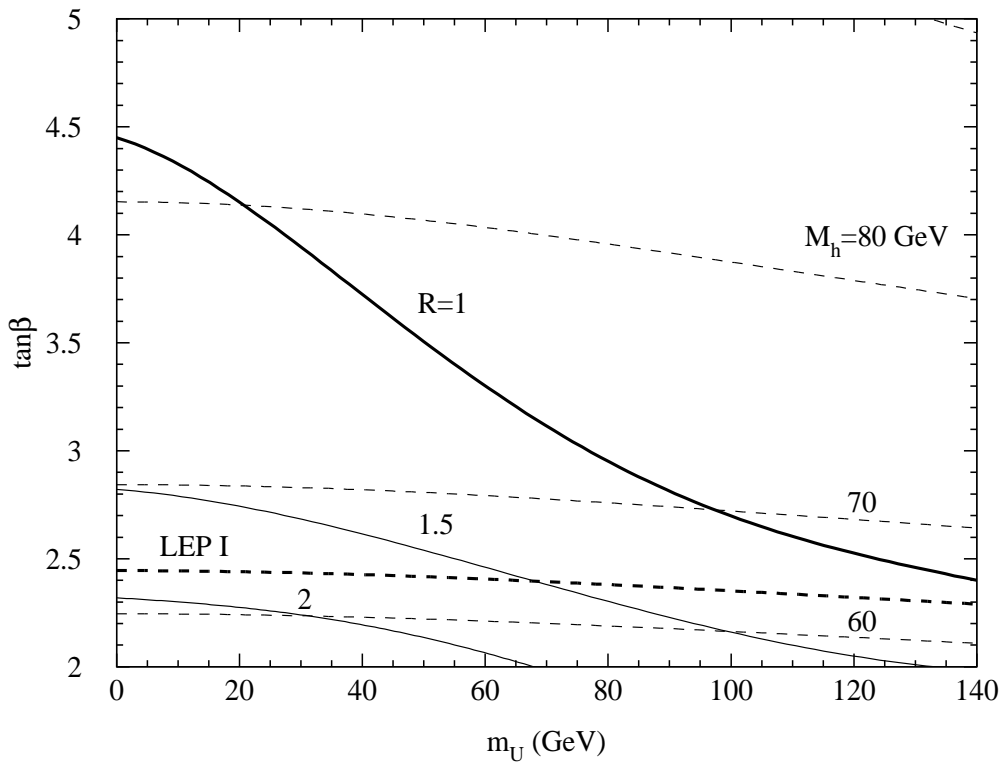,angle=-90,height=10cm,width=12cm,bbllx=2.cm,bblly=4.cm,bburx=19.cm,bbury=25cm}}
\caption{\footnotesize Contour lines of ${\cal R}$ (solid) and $m_h$ (dashed) in the 
$(m_{U},\tan\beta)$ plane for $M_t=156\ GeV$ and $m_Q=70\ GeV$.} 
\end{figure} 

In the previous numerical example we have chosen the parameters to 
maximize the effect on ${\cal R}$. We explore now how ${\cal R}$ changes 
when we move in parameter space.  
Figure 3 is a contour plot of ${\cal R}$ (solid lines) in the 
plane
($m_{U},\tan\beta$) for $m_t=150\ GeV$ ($M_t=156\ GeV$) and $m_{Q}=70\ 
GeV$. The dashed lines give the Higgs boson mass\footnote{To compute 
the Higgs mass accurately is important to include loop corrections which 
can be sizeable \cite{Radmh,Radmh2}.} $m_h$ in GeV. Thick lines are used to 
single 
out the lines ${\cal R}=1$ and $m_h=64\ GeV$. The region where electroweak 
baryogenesis could in principle be viable and the LEP I bound on $m_h$ 
evaded is now sizeable. Values of $\tan\beta$ larger than in the 
one-loop case are allowed and Higgs masses up to $\sim 80 \ GeV$ can be 
reached.
The decrease of ${\cal R}$ with increasing $m_U$ is just the effect of a 
growing screening in $\otr$. 
For large values of $m_U$ the high temperature expansion breaks down 
and one should evaluate two-loop contributions numerically. We cut the 
x-axis for $m_U=140\ GeV$ where the high T expansion is still safe.
\begin{figure}[hbt]
\centerline{
\psfig{figure=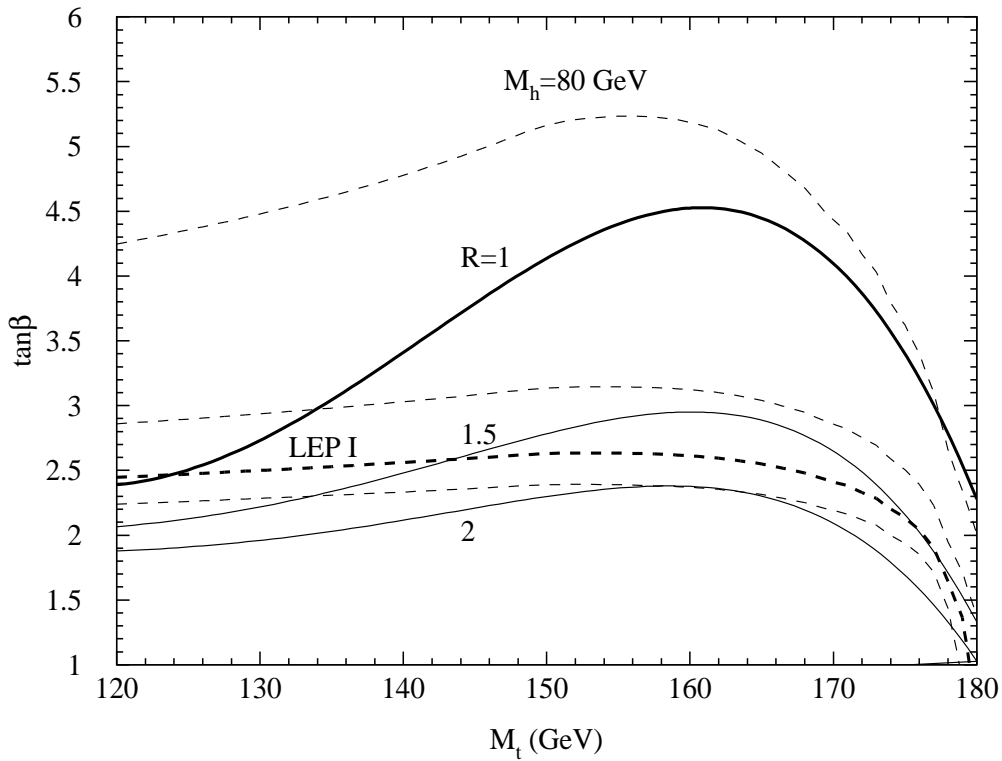,angle=-90,height=10cm,width=12cm,bbllx=2.cm,bblly=4.cm,bburx=19.cm,bbury=25cm}}
\caption{\footnotesize Contour lines of ${\cal R}$ (solid) and $m_h$ (dashed) in the 
$(M_t,\tan\beta)$ plane for $m_U=0$ and $m_Q$ above the limit to avoid 
the $\Delta\rho$ constraint.} 
\end{figure} 

Finally, figure 4 shows contour lines of constant ${\cal R}$ and $m_h$ in 
the plane
$(M_t,\tan\beta)$. Thick solid lines for ${\cal R}=1$ and $m_h=64\ GeV$ 
delimit 
a "baryogenesis hole" where the electroweak phase transition is strong and
the lightest Higgs mass is above the LEP I limit. The region corresponds 
to $125\ GeV \leq M_t \leq 185\ GeV$ and $\tan\beta \leq 4.5$. 
In this plot we have set
$m_{U}=0$ and $m_{Q}$ has been fixed at its lowest value such that 
the constraint on $\Delta\rho$ is evaded (fixing for simplicity 
$\tan\beta=2$ to compute this minimum of $m_{Q}$). The range of $m_{Q}$
then varies from zero for $M_t=120\ GeV$ to order $1\ TeV$ for $M_t=180\ 
GeV$. 
The dependence of ${\cal R}$ contours on $M_t$ is then easy to understand: 
for 
low values of $M_t$ the effect of squarks is smaller because the Yukawa 
coupling is smaller. In this region, increasing $M_t$ one gets larger 
values of ${\cal R}$. For larger $M_t$, two effects collaborate to change 
this 
tendency. Large $M_t$ requires large $m_{Q}$ to avoid problems with 
$\Delta\rho$ so that left-handed squarks are required to be heavy. In turn, 
heavy 
squarks give a sizeable radiative correction to the Higgs mass making it
heavier and correspondingly decreasing the strength of the transition. 
This is purely a zero temperature effect. For the finite temperature
corrections, heavy left-handed squarks are in fact decoupled and do not
contribute: ${\cal R}$ in this region has been calculated with 
contributions of
right handed squarks only. Therefore, the effect is reduced and ${\cal R}$ 
drops. The net effect is that one ends up with a closed region in the 
$(M_t,\tan\beta)$ plane where electroweak baryogenesis may be successful.
It is interesting to note that this region requires large $M_t$ in a 
range compatible with the CDF-D0 range. Furthermore $m_h$ is bounded to 
be below $\sim 80\ GeV$ so that LEP II will scan this "baryogenesis hole" in 
the near future testing then one of the crucial ingredients for the 
viability of electroweak baryogenesis in the MSSM. 

\section{Conclusion}
In the MSSM light stops can dominate completely the electroweak phase 
transition in the early Universe. We have shown that the strength of this 
transition is dramatically affected by QCD corrections that first appear
at two-loop order. They make the transition strongly first order in some
non marginal region of parameter space compatible with experimental 
constraints. This is in sharp contrast with the situation in the 
Standard Model where transverse gauge bosons drive the transition which 
is weak for realistic Higgs masses.
  
The allowed region found corresponds to $\tan\beta$ small within a 
reasonable 
range ($\tan\beta \simlt 4.5$), squarks as light as allowed by 
electroweak 
precision constraints and only one light 
Higgs boson, with 
Standard couplings and mass as large as $80 \ GeV$. Such Higgs scalar
is within the reach of LEP II so that these region will be explored
in the forthcoming years.

The sizeable increase in the strength of the electroweak phase transition
can be decisive to open the window for electroweak baryogenesis in the 
MSSM and should encourage further studies of many different aspects of the 
problem \cite{hune} still not well understood and necessary towards a 
complete mechanism of electroweak (supersymmetric) baryogenesis.
\vspace{0.5cm}

{\bf Note Added}

In a very recent paper \cite{cqw} the electroweak phase transition in the
MSSM is studied in the previously unexplored case $m_U^2<0$ which is 
shown to be phenomenologically allowed. In such region of parameter space
the screening of $\otr$ is then reduced and ${\cal R}$ can be larger than
1 even at one-loop. Then the allowed areas found in the present paper can be
continued into the $m_U^2<0$ region improving further the prospects for 
supersymmetric electroweak baryogenesis.
\vspace{0.5cm}

{\bf Acknowledgements}

Thanks go to A.~Brignole, W.~Buchm\"uller, M.~Cacciari, M.~Carena, 
A.~Hebecker, M.Qui- r\'os, J.Rosiek, M.~Spira, C.~Wagner and F.~Zwirner 
for useful comments, discussions, and suggestions regarding different 
aspects of this work.
\vspace{0.5cm}

\section*{ Appendix A}
\vspace{0.5cm}
 
We reproduce here the leading parts of the T-dependent self energies for 
the different species of bosonic particles. Most of these results were
presented in ref.~\cite{I}. Some errata are corrected now (see 
Ref.\cite{coes} for details).
It is assumed that $m_A$ is large so that only one light Higgs doublet 
is present at the electroweak scale. The second doublet is decoupled
and does not contribute to the thermal polarizations written below. As 
in ref.~\cite{I} terms outside curly brackets come from Standard Model 
contributions and third generation squarks. The rest of the supersymmetric
contributions is inside the brackets and is written for completeness.

Note that, at leading order, these thermal masses are equal for particles 
inside the same $SU(2)$ multiplet.  They are given by 
\bear
\Pi_{W_L} &=& \frac{7}{3} g^2 T^2  +
\left\{ 2 g^2 T^2 \right\} \nonumber\, ,\\
\Pi_{B_L} &=& \frac{22}{9} g'\,^2 T^2 +
\left\{ \frac{26}{9} g'\,^2 T^2 \right\} \nonumber\, , \\
\Pi_h &=& \Pi_{\chi} =
\frac{1}{16}  (g^2+g'\,^2) \cos^22\beta T^2
+ \frac{3}{16} g^2 T^2 + \frac{1}{16} g'\,^2 T^2 +
\frac{3}{4} h_t^2\sin^2\beta  T^2\nonumber\,  \\
&+& 
\left\{\frac{1}{8} g^2 T^2 + \frac{1}{24} g'\,^2 T^2 \right\}\nonumber ,\\
\Pi_{\tilde{t}_L} &=&\Pi_{\tilde{b}_L} = \frac{4}{9} g_s^2 T^2 + 
\frac{1}{4} g^2 T^2 + \frac{1}{216} g'\,^2(2-3\cos 2\beta) T^2 +
\frac{1}{12} h_t^2 (1 + \sin^2\beta) T^2\nonumber\,  \\
&+&\left\{\frac{2}{9} g_s^2 T^2 + \frac{1}{8} g^2 T^2
+ \frac{1}{216} g'\,^2 T^2 + \frac{1}{12}h_t^2T^2\right\} \nonumber\, ,\\
\Pi_{\tilde{t}_R} &=& \frac{4}{9} g_s^2 T^2 + \frac{1}{54} g'\,^2(8+3\cos
2\beta) T^2 + \frac{1}{6} h_t^2 (1 + \sin^2\beta) T^2 \nonumber\,  \\
&+&\left\{\frac{2}{9} g_s^2 T^2 + \frac{2}{27} g'\,^2 T^2 
+ \frac{1}{6}h_t^2T^2\right\} \nonumber\, ,\\
\Pi_{\tilde{b}_R} &=& \frac{4}{9} g_s^2 T^2 + \frac{1}{108} g'\,^2(4-3\cos 
2\beta) T^2 +
\left\{\frac{2}{9} g_s^2 T^2 + \frac{1}{54} g'\,^2 T^2 \right\} \nonumber\, .
\eear

We complete these formulae giving the polarization for longitudinal 
gluons 
\[
\Pi_{g_L}=\frac{8}{3}g_s^2T^2+\left\{\frac{11}{6}g_s^2T^2
\right\}\nonumber.
\]
This last result is needed for a two-loop calculation of the effective 
potential.
\vspace{0.5cm}

\section*{ Appendix B}
\vspace{0.5cm}

We write here the resummed supersymmetric two-loop corrections to the finite 
temperature effective potential in the MSSM framework described in the text.
In particular we consider $g'=h_b=0$ so that in the formulas below we
make no distinction between $m_Z$ and $m_W$ which are called $M$. We assume
also that left-right mixing in the stop sector is negligible.
Our notation is that of Arnold and Espinosa in \cite{TWOLSM} and the 
labelling of different contributions corresponds to our figure 1. As usual,
$N_c=3$ counts the number of colours.
\bear
V_{(a')}&=&-\frac{g^2}{8}N_c\left[{\cal D}_{SSV}(\mtl,\mtl,M)+
{\cal D}_{SSV}(\mbl,\mbl,M)+
4 {\cal D}_{SSV}(\mtl,\mbl,M)
\right]\nonumber\\
&-&\frac{g_s^2}{4}(N_c^2-1)\left[{\cal D}_{SSV}(\mtl,\mtl,0)+
{\cal D}_{SSV}(\mbl,\mbl,0)\right.\nonumber\\
&+&\left.{\cal D}_{SSV}(\mtr,\mtr,0)+
{\cal D}_{SSV}(\mbr,\mbr,0)
\right],\nonumber\\
V_{(p')}&=&-\left[\left(
h_t^2\sin^2\beta+\frac{g^2}{4}\cos2\beta\right)^2 {\overline 
{\mathrm H}}(m_h,\mtl,\mtl)+
\left(\frac{g^2}{4}\cos2\beta\right)^2 {\overline {\mathrm 
H}}(m_h,\mbl,\mbl)\right. \nonumber\\
&+&\left.(h_t^2\sin^2\beta)^2 {\overline {\mathrm H}}(m_h,\mtr,\mtr)+
\left(
h_t^2\sin^2\beta+\frac{1}{2}g^2\cos2\beta\right)^2 
{\overline {\mathrm H}}(m_{\chi},\mtl,\mbl) \right]\frac{\varphi^2}{2}N_c
,\nonumber\\
V_{(z'_{1,2})}&=&-\frac{1}{4}g_s^2(N_c^2-1)\left[{\cal D}_{SV}(\mtl,0)
+{\cal D}_{SV}(\mbl,0)+{\cal D}_{SV}(\mtr,0)+{\cal D}_{SV}(\mbr,0)
\right]\nonumber\\
&-&\frac{3}{8} g^2 N_c\left[{\cal D}_{SV}(\mtl,M)+
{\cal D}_{SV}(\mbl,M)
\right],\nonumber\\
V_{(z'_{3,4})}&=&
\frac{g^2}{4}N_c(2-N_c){\cal D}_{S}(\mtl,\mbl)
+h_t^2N_c\left[{\cal D}_{S}(\mtl,\mtr)+{\cal 
D}_{S}(\mbl,\mtr)\right]\nonumber\\
&+&\left(\frac{g^2}{8}+\frac{g_s^2}{6}\right)N_c(N_c+1)\left[
{\cal D}_{S}(\mtl,\mtl)+{\cal D}_{S}(\mbl,\mbl)
\right]\nonumber\\
&+&\frac{g_s^2}{6}N_c(N_c+1)\left[{\cal D}_{S}(\mtr,\mtr)+{\cal 
D}_{S}(\mbr,\mbr) \right]\nonumber\\
&+&N_c\left(\frac{1}{2}h_t^2\sin^2\beta+\frac{1}{8}g^2\cos2\beta\right)\left[
{\cal D}_{S}(\mtl,m_h)+2{\cal D}_{S}(\mbl,m_{\chi})+{\cal 
D}_{S}(\mtl,m_{\chi}) \right]\nonumber\\
&-&\frac{1}{8}N_cg^2\cos2\beta\left[
{\cal D}_{S}(\mbl,m_h)+2{\cal D}_{S}(\mtl,m_{\chi})+{\cal 
D}_{S}(\mbl,m_{\chi}) \right]\nonumber\\
&+&\frac{1}{2}N_ch_t^2\sin^2\beta\left[
{\cal D}_{S}(\mtr,m_h)+2{\cal D}_{S}(\mtr,m_{\chi})+{\cal 
D}_{S}(\mtr,m_{\chi}) \right],\nonumber
\eear
where ${\cal D}_{S}(m_1,m_2)$ is simply the resummed version of
$$
{\cal D}^{
\hbox{\rlap{$\scriptstyle R$} {\hspace*{-1mm}$/$}}
}_{S}(m_1,m_2)=I(m_1)I(m_2). 
$$

In the previous formulae, dimensional regularization (with 
$n-1=3-2\epsilon$) is used to evaluate divergent integrals. Poles in 
$1/\epsilon$ and $\iota_\epsilon$-dependent terms cancel when 
counterterms are included. In addition, counterterms contribute a finite 
piece to the potential \cite{TWOLSM}. To compute the finite piece 
due to third generation squarks note that the counterterm potential can 
be written now as
\def\dzm{\delta  Z_{\mu^2}}
\def\dzp{\delta Z_{\varphi^2}}
\def\dzg{\delta Z_{g^2}}
\def\dzgp{\delta Z_{g'^2}}
\def\dzht{\delta Z_{h_{t,SM}^2}}
\def\dzl{\delta Z_{\lambda}}
\bear
\nonumber
\delta V 
&=&\frac{1}{2}\varphi^2\left\{-\mu^2(\dzm+\dzp)+I^{\epsilon}_{\beta}
\left[6\lambda(\dzl+
\dzp)+6 h_{t,SM}^2(\dzht+\dzp)
\right.
\right.\nonumber\\
&+&\left.\left.
\frac{1}{4}(3-2\epsilon)\left(3g^2(\dzg+\dzp)+g'^2(\dzgp+\dzp)\right)\right]
-6h_{t,SM}^2I^\epsilon_{ f\beta}(\dzht+\dzp)\right\}\nonumber\\
&+&\frac{\lambda}{8}\varphi^4(\dzl+\dzp),
\eear
with 
\bear
\nonumber
I^\epsilon_\beta=\frac{T^2}{12}(1+\epsilon\iota_\epsilon)&,&
I^\epsilon_{ f\beta}=-\frac{T^2}{24}[1+\epsilon(\iota_\epsilon-\log4)].
\eear

The $\ov{\mathrm MS}$ renormalization functions $\delta Z$ have no 
finite part, so the finite ($\iota_\epsilon$-independent) piece of $\delta V$
is given by the $(3-2\epsilon)$ and $I^\epsilon_{f\beta}$ terms.
The required contributions of third generation squarks to the 
renormalization functions are
\be
\nonumber
\delta_{\tilde{q}_3} 
Z_{g^2}=\frac{1}{2}g^2\frac{1}{16\pi^2\epsilon},\;\; \;\;
\delta_{\tilde{q}_3} 
Z_{g'^2}=\frac{11}{18}g'^2\frac{1}{16\pi^2\epsilon}, \;\;\;\;
\delta_{\tilde{q}_3}Z_{\varphi^2}=\delta_{\tilde{q}_3} 
Z_{h_{t,SM}^2}=0. \ee
The third equation follows from the fact that squarks are scalars while 
the fourth is a consequence of our assumptions about the supersymmetric 
spectrum, that has no light ${\mathrm R}~=~-1$ fermions. The finite 
contribution is then
\be
\nonumber
\delta 
V=-\frac{T^2}{16\pi^2}\frac{\varphi^2}{96}\left(3g^4+\frac{11}{9}g'^4\right).
\ee
\vspace{0.5cm}

\section*{ Appendix C}
\vspace{0.5cm}

To include the effect of the left-right squark mixing
appropriate mixing angles, $\alpha_t$ for stops and $\alpha_b$ for
sbottoms, should be defined. When temperature corrections are included
in the mass matrix for the zero Matsubara modes, the corresponding
mixing angles will turn out to be temperature dependent. This effect
has to be taken into account when writing the resummed 
effective potential and zero modes should be treated independently. It is 
then possible to generalize the previous expression for the effective 
potential writing it in terms of unresummed ${\cal D}$'s plus extra
terms for the resummation of zero modes. A similar situation arises with the 
photon-Z mixing in the Standard Model. 
In the same way that this complication is circumvented in the Standard Model
just by setting $g'=0$, a reasonable assumption for the squark case would
be to approximate the squark thermal masses by its universal QCD 
contribution 
$4g_s^2T^2/9$ and neglect the rest. In this way the thermal correction to the
squared mass matrices would be diagonal and the mixing angles would be 
temperature independent. In this approximation, the generalization of the 
previous expression for the potential is straightforward.
Writing
\be
\left\{\begin{array}{cl}
\tilde{t}_1&=\cos\alpha_t \tilde{t}_L +\sin\alpha_t \tilde{t}_R\\
\tilde{t}_2&=\cos\alpha_t \tilde{t}_R -\sin\alpha_t \tilde{t}_L 
\end{array}\right.\;\;\;\;\;\;
\left\{\begin{array}{cl}
\tilde{b}_1&=\cos\alpha_b \tilde{b}_L +\sin\alpha_b \tilde{b}_R\\
\tilde{b}_2&=\cos\alpha_b \tilde{b}_R -\sin\alpha_b \tilde{b}_L 
\end{array}\right.
\ee
we obtain (using the abbreviations $c_t=\cos\alpha_t$, 
$c_{2\beta}=\cos2\beta$, etc) \bear
V_{(a')}&=&-\frac{g^2}{8}N_c\left[c_t^4{\cal 
D}_{SSV}(\mstu,\mstu,M)+ s_t^4{\cal D}_{SSV}(\mstd,\mstd,M)+
2c_t^2s_t^2 {\cal D}_{SSV}(\mstu,\mstd,M)
\right]\nonumber\\
&-&\frac{g^2}{8}N_c\left[c_b^4{\cal D}_{SSV}(\msbu,\msbu,M)+
s_b^4{\cal D}_{SSV}(\msbd,\msbd,M)+
2c_b^2s_b^2 {\cal D}_{SSV}(\msbu,\msbd,M)
\right]\nonumber\\
&-&\frac{g^2}{2}N_c\left[c_t^2c_b^2{\cal D}_{SSV}(\mstu,\msbu,M)+
s_t^2s_b^2{\cal D}_{SSV}(\mstd,\msbd,M)+
c_t^2s_b^2{\cal D}_{SSV}(\mstu,\msbd,M)\right.\nonumber\\
&+&\left. s_t^2c_b^2{\cal D}_{SSV}(\mstd,\msbu,M)
\right]
-\frac{g_s^2}{4}(N_c^2-1)\left[{\cal D}_{SSV}(\mstu,\mstu,0)+
{\cal D}_{SSV}(\msbu,\msbu,0)\right.\nonumber\\
&+&\left.
{\cal D}_{SSV}(\mstd,\mstd,0)
+{\cal D}_{SSV}(\msbd,\msbd,0)
\right],\nonumber
\eear
\bear
V_{(p')}&=&-\frac{N_c}{2}\left\{
2\left[
\frac{g^2}{4}\varphi c_ts_tc_{2\beta} 
+\frac{h_t}{\sqrt{2}}(A_t s_{\beta}+\mu c_{\beta})c_{2t}\right]^2 
{\overline {\mathrm H}}(m_h,\mstu,\mstd)\right.
\nonumber\\
&+&
\left[
(h_t^2 s_{\beta}^2+\frac{g^2}{4}c_t^2c_{2\beta})\varphi 
-h_t\sqrt{2}(A_t s_{\beta}+\mu c_{\beta})c_ts_t\right]^2 {\overline {\mathrm 
H}}(m_h,\mstu,\mstu)\nonumber\\
&+&\left.\left[
(h_t^2 s_{\beta}^2+\frac{g^2}{4}s_t^2 c_{2\beta})\varphi 
+h_t\sqrt{2}(A_t s_{\beta}+\mu c_{\beta})c_ts_t\right]^2 {\overline {\mathrm 
H}}(m_h,\mstd,\mstd)\right\}\nonumber\\
&-&\frac{N_c}{2}\left\{
2\left[
\frac{g^2}{4}\varphi c_bs_bc_{2\beta} 
+\frac{h_b}{\sqrt{2}}(A_b c_{\beta}+\mu s_{\beta})c_{2b}\right]^2 
{\overline {\mathrm H}}(m_h,\msbu,\msbd)\right.
\nonumber\\
&+&
\left[
\frac{g^2}{4}\varphi c_b^2c_{2\beta} 
-h_b\sqrt{2}(A_b c_{\beta}+\mu s_{\beta})c_bs_b\right]^2 {\overline {\mathrm 
H}}(m_h,\msbu,\msbu)\nonumber\\
&+&\left.\left[
\frac{g^2}{4}\varphi s_b^2 c_{2\beta} 
+h_b\sqrt{2}(A_b c_{\beta}+\mu s_{\beta})c_bs_b\right]^2 {\overline {\mathrm 
H}}(m_h,\msbd,\msbd)\right\}\nonumber\\
&-&\frac{N_c}{2}\varphi^2
\left(\frac{g^2}{4} c_{2\beta}\right)^2\left[c_b^4 {\overline {\mathrm 
H}}(m_h,\msbu,\msbu)+s_b^4 {\overline {\mathrm 
H}}(m_h,\msbd,\msbd)+2c_b^2s_b^2 {\overline {\mathrm 
H}}(m_h,\msbu,\msbd)
\right]\nonumber\\
&-&\frac{N_c}{2}\left[h_t^2(A_t s_{\beta}+\mu c_{\beta})^2{\overline 
{\mathrm H}}(m_{\chi},\mstu,\mstd)+
h_b^2(A_b c_{\beta}+\mu s_{\beta})^2{\overline {\mathrm 
H}}(m_{\chi},\msbu,\msbd)\right]\nonumber\\
&-&N_c\left\{\left[
\frac{\varphi}{\sqrt{2}}(h_t^2 s^2_{\beta}+\frac{g^2}{2} c_{2\beta})c_tc_b
+h_tX_ts_tc_b
-h_bX_bc_ts_b\right]^2 {\overline {\mathrm 
H}}(m_{\chi},\mstu,\msbu)\right.\nonumber\\
&+&\left[
\frac{\varphi}{\sqrt{2}}(h_t^2 s^2_{\beta}+\frac{g^2}{2} c_{2\beta})c_ts_b
+h_tX_ts_ts_b
+h_bX_bc_tc_b\right]^2 {\overline {\mathrm 
H}}(m_{\chi},\mstu,\msbd)\nonumber\\
&+&
\left[
\frac{\varphi}{\sqrt{2}}(h_t^2 s^2_{\beta}+\frac{g^2}{2} c_{2\beta})s_tc_b
-h_tX_tc_tc_b
-h_bX_bs_ts_b\right]^2 {\overline {\mathrm 
H}}(m_{\chi},\mstd,\msbu)\nonumber\\
&+&
\left.
\left[
\frac{\varphi}{\sqrt{2}}(h_t^2 s^2_{\beta}+\frac{g^2}{2}c_{2\beta})s_ts_b
-h_tX_tc_ts_b
+h_bX_bs_tc_b\right]^2 {\overline {\mathrm 
H}}(m_{\chi},\mstd,\msbd)
\right\},\nonumber
\eear
where we keep $h_b$ only when it appears in combination with $X_b$.
\bear
V_{(z'_{1,2})}&=&-\frac{1}{4}g_s^2(N_c^2-1)\left[{\cal 
D}_{SV}(\mstu,0)
+{\cal D}_{SV}(\msbu,0)+{\cal D}_{SV}(\mstd,0)+{\cal D}_{SV}(\msbd,0)
\right]\nonumber\\
&-&\frac{3}{8} g^2 N_c\left[c_t^2{\cal D}_{SV}(\mstu,M)+
s_t^2{\cal D}_{SV}(\mstd,M)+
c_b^2{\cal D}_{SV}(\msbu,M)+
s_b^2{\cal D}_{SV}(\msbd,M)\right],\nonumber
\eear
\bear
V_{(z'_{3,4})}&=&\frac{g_s^2}{6}N_c(N_c+1)\left[
{\cal D}_{S}(\mstu,\mstu)+{\cal D}_{S}(\msbu,\msbu)+
{\cal D}_{S}(\mstd,\mstd)  
\right.\nonumber\\
&+& \left.{\cal D}_{S}(\msbd,\msbd)\right]+\frac{N_c}{2}\left[
\left(h_t^2 s^2_{\beta}+\frac{g^2}{4}c_t^2 c_{2\beta}\right)
[{\cal D}_{S}(m_h,\mstu)+{\cal D}_{S}(m_{\chi},\mstu)]\right.
\nonumber\\
&+&\left.
\left(h_t^2 s^2_{\beta}+\frac{g^2}{4}s_t^2 c_{2\beta}\right)
[{\cal D}_{S}(m_h,\mstd)+{\cal D}_{S}(m_{\chi},\mstd)]
\right]+N_ch_t^2\left[
c_b^2s_t^2{\cal D}_{S}(\msbu,\mstu)\right.\nonumber\\
&+&\left.
c_b^2c_t^2{\cal D}_{S}(\msbu,\mstd)+
s_b^2s_t^2{\cal D}_{S}(\msbd,\mstu)+
s_b^2c_t^2{\cal D}_{S}(\msbd,\mstd)
\right]\nonumber\\
&-&\frac{N_cg^2 c_{2\beta}}{8}
\left[c_b^2[{\cal D}_{S}(m_h,\msbu)-{\cal D}_{S}(m_{\chi},\msbu)]+
s_b^2[{\cal D}_{S}(m_h,\msbd)-{\cal D}_{S}(m_{\chi},\msbd)]
\right.\nonumber\\
&+&\left.
2 c_t^2{\cal D}_{S}(m_{\chi},\mstu)
+2 s_t^2{\cal D}_{S}(m_{\chi},\mstd)\right]\nonumber\\
&+&N_ch_t^2 s^2_{\beta}
\left[
s_t^2{\cal D}_{S}(m_{\chi},\mstu)
+c_t^2{\cal D}_{S}(m_{\chi},\mstd)
+c_b^2{\cal D}_{S}(m_{\chi},\msbu)
+s_b^2{\cal D}_{S}(m_{\chi},\msbd)\right]\nonumber\\
&+&N_ch_t^2c_t^2s_t^2\left[(N_c+1)\left[{\cal D}_{S}(\mstu,\mstu)+
{\cal D}_{S}(\mstd,\mstd)\right]-2N_cc_t^2s_t^2
{\cal D}_{S}(\mstu,\mstd)\right]\nonumber\\
&+&\frac{g^2}{4}N_c(2-N_c)\left[c_t^2c_b^2{\cal 
D}_{S}(\mstu,\msbu)
+c_t^2s_b^2{\cal D}_{S}(\mstu,\msbd)
+s_t^2c_b^2{\cal D}_{S}(\mstd,\msbu)\right.\nonumber\\
&+&\left. s_t^2s_b^2{\cal D}_{S}(\mstd,\msbd)\right] 
+N_ch_t^2(c_t^4+s_t^4){\cal D}_{S}(\mstu,\mstd)\nonumber\\
&+&\frac{1}{8}g^2N_c(N_c+1)\left[
c_t^4{\cal D}_{S}(\mstu,\mstu)+
2c_t^2s_t^2{\cal D}_{S}(\mstu,\mstd)+
s_t^4{\cal D}_{S}(\mstd,\mstd)
\right]\nonumber\\
&+&\frac{1}{8}g^2N_c(N_c+1)\left[
c_b^4{\cal D}_{S}(\msbu,\msbu)+
2c_b^2s_b^2{\cal D}_{S}(\msbu,\msbd)+
s_b^4{\cal D}_{S}(\msbd,\msbd)
\right].\nonumber
\eear

\end{document}